\pdfoutput=1
\documentclass[10pt,twocolumn,english,aps,prb,superscriptaddress,amsmath,amssymb,floatfix,footinbib]{revtex4-2}
\usepackage[english]{babel}
\usepackage{amsmath,amsfonts,amssymb}
\usepackage[T1]{fontenc}
\usepackage[utf8]{inputenc} 
\usepackage{lmodern} 
\usepackage{xurl}
\usepackage{textcomp}
\usepackage{amsmath}
\usepackage{siunitx}
\usepackage[version=4]{mhchem}
\usepackage{amsfonts}
\usepackage{amssymb}

\usepackage{epstopdf}
\usepackage{graphicx}
\graphicspath{{./}}

\usepackage[colorlinks=true,citecolor=blue,linkcolor=magenta]{hyperref}

\newcommand{\fref}[1]{Fig.~\ref{#1}}
\newcommand{\SiN}[0]{$\mathrm{Si}_3\mathrm{N}_4~$}
\newcommand{\LiNbO}{\ensuremath{\mathrm{LiNbO}_3}}
\newcommand{\LiTaO}{\ensuremath{\mathrm{LiTaO}_3}}
\newcommand{\SiO}[0]{$\mathrm{SiO}_2~$ }

\newcommand{\um}{\text{\textmu m}}

\makeatletter
\def\bibsection{%
  \par
  \begingroup
   \baselineskip26\p@
   \bib@device{\hsize}{72\p@}%
  \endgroup
  \nobreak\@nobreaktrue
  \addvspace{19\p@}%
}%
\makeatother

\usepackage{environ}
\newif\ifarxivlayout
\arxivlayouttrue
\newcount\paperfigureindex
\makeatletter
\newcommand{\arxivfigurepage}[1]{%
  \begin{figure*}[p]
    \noindent\begin{minipage}[t][\textheight][s]{\textwidth}%
      \def\@captype{figure}%
      #1\par
    \end{minipage}%
  \end{figure*}%
}
\NewEnviron{paperfigure}{%
  \global\advance\paperfigureindex by 1
  \ifarxivlayout
    \ifnum\paperfigureindex=1
      \arxivfigurepage{\BODY}%
    \else
      \begin{figure*}[!t]\BODY\end{figure*}%
    \fi
  \else
    \begin{figure*}[!t]\BODY\end{figure*}%
  \fi
}
\makeatother
\newcommand{\paperfiguregraphic}[2][]{%
  \ifarxivlayout
    \ifnum\paperfigureindex=1
      \includegraphics[width=\linewidth,height=0.82\textheight,keepaspectratio]{#2.pdf}%
      \par\vfill
    \else
      \includegraphics[width=\linewidth,#1]{#2.pdf}%
    \fi
  \else
    \includegraphics[width=\linewidth,#1]{#2.pdf}%
  \fi
}

\makeatletter
\def\doauthor#1#2#3{%
  \ignorespaces#1\unskip
  \begingroup
   #3%
  \@if@empty{#2}{\endgroup{}{}}{\endgroup{\comma@space}{}\frontmatter@footnote{#2}}%
  \@listcomma
  \space \@listand
}%
\makeatother

\begin{document}
		\title{Terabit-scale wavelength-multiplexed transmitter\texorpdfstring{\\}{ }enabled by thin-film lithium tantalate}

	\author{Shivaprasad U. Hulyal}
	\thanks{These authors contributed equally to this work.}
	\affiliation{Institute of Physics, Swiss Federal Institute of Technology Lausanne (EPFL), CH-1015 Lausanne, Switzerland}
	\affiliation{Institute of Electrical and Micro Engineering, Swiss Federal Institute of Technology Lausanne (EPFL), CH-1015 Lausanne, Switzerland}

	\author{Dengyang Fang}
	\thanks{These authors contributed equally to this work.}
	\affiliation{Institute of Photonics and Quantum Electronics (IPQ), Karlsruhe Institute of Technology (KIT), 76131 Karlsruhe, Germany}

	\author{Jiachen Cai}
	\thanks{These authors contributed equally to this work.}
	\affiliation{Institute of Physics, Swiss Federal Institute of Technology Lausanne (EPFL), CH-1015 Lausanne, Switzerland}
	\affiliation{Institute of Electrical and Micro Engineering, Swiss Federal Institute of Technology Lausanne (EPFL), CH-1015 Lausanne, Switzerland}

	\author{Xinru Ji}
	\affiliation{Institute of Physics, Swiss Federal Institute of Technology Lausanne (EPFL), CH-1015 Lausanne, Switzerland}
	\affiliation{Institute of Electrical and Micro Engineering, Swiss Federal Institute of Technology Lausanne (EPFL), CH-1015 Lausanne, Switzerland}

	\author{Alexander Kotz}
	\affiliation{Institute of Photonics and Quantum Electronics (IPQ), Karlsruhe Institute of Technology (KIT), 76131 Karlsruhe, Germany}

	\author{Jiale Sun}
	\affiliation{Institute of Physics, Swiss Federal Institute of Technology Lausanne (EPFL), CH-1015 Lausanne, Switzerland}
	\affiliation{Institute of Electrical and Micro Engineering, Swiss Federal Institute of Technology Lausanne (EPFL), CH-1015 Lausanne, Switzerland}

	\author{Christian Koos}
	\email[]{christian.koos@kit.edu}
	\affiliation{Institute of Photonics and Quantum Electronics (IPQ), Karlsruhe Institute of Technology (KIT), 76131 Karlsruhe, Germany}

	\author{Tobias J. Kippenberg}
	\email[]{tobias.kippenberg@epfl.ch}
	\affiliation{Institute of Physics, Swiss Federal Institute of Technology Lausanne (EPFL), CH-1015 Lausanne, Switzerland}
	\affiliation{Institute of Electrical and Micro Engineering, Swiss Federal Institute of Technology Lausanne (EPFL), CH-1015 Lausanne, Switzerland}

	\medskip
	\maketitle

	\textbf{\noindent
		Artificial-intelligence systems face a growing data-movement bottleneck and require higher interconnect bandwidth without a proportional increase in electrical power consumption or the number of laser sources and fibre connections. Moving optical interfaces closer to computing chips reduces electrical transmission loss, while demanding optical engines that combine high channel capacity, broadband modulation and simple packaging. Although wavelength multiplexing relieves pressure on the fibre budget, integrating shared multi-wavelength sources with dense wavelength routing and broadband modulation in a compact format remains challenging.
		Here we demonstrate a monolithic transmitter that brings these functions together on a single material platform, enabled by the low birefringence and native Pockels response of thin-film lithium tantalate. The device separates, independently modulates and recombines eight wavelength channels between one optical input and output. This architecture retains a two-fibre interface as the channel count grows, directly addressing the packaging burden of parallel optical links.
		Measurements across the eight channels establish an aggregate line rate of 1.6~Tbit~s$^{-1}$. Operation with a shared optical frequency comb achieves an estimated net data rate of 0.58~Tbit~s$^{-1}$ across four channels, demonstrating compatibility with a common multi-wavelength source.
		Scaling this architecture into multiple wavelength-multiplexed engines offers a path towards 10~Tbit~s$^{-1}$ optical interfaces with high bandwidth per unit package edge, positioning thin-film lithium tantalate for the next generation of co-packaged optical interconnects.}

	\newpage 
    \section*{Introduction}
    \noindent
	Scaling artificial-intelligence systems requires optical interfaces that deliver greater interconnect capacity within tight constraints on power consumption, footprint and fibre count. Training and inference require large amounts of data to move between accelerators, memory stacks and switches. System performance therefore depends on both the computational throughput of individual chips and the bandwidth, latency and energy cost of the interconnect fabric~\cite{kim_co-packaged_2026,tan_co-packaged_2023}. As transmission distances and signalling rates increase, microwave propagation loss in electrical interconnects raises the power needed to maintain signal integrity and limits achievable bandwidth density. Optical interconnects offer low propagation loss and ultrahigh bandwidth, motivating their use throughout the computing architecture as well as between racks~\cite{miller_attojoule_2017}.

	Realizing these benefits also requires shortening the electrical path to the optical interface. With front-panel pluggable transceivers, high-speed signals still traverse centimetre-scale copper traces between the application-specific integrated circuit (ASIC) and the optical module~\cite{kim_co-packaged_2026}. Co-packaged optics (CPO) shortens this path by integrating photonic and electronic circuits within a common package, using 2.5D interposers or three-dimensional stacking. This approach is also being developed for graphics processing units (GPUs) and other AI accelerators through the integration of optical I/O chiplets.

	\begin{paperfigure}
		\centering
		\paperfiguregraphic[height=0.76\textheight,keepaspectratio]{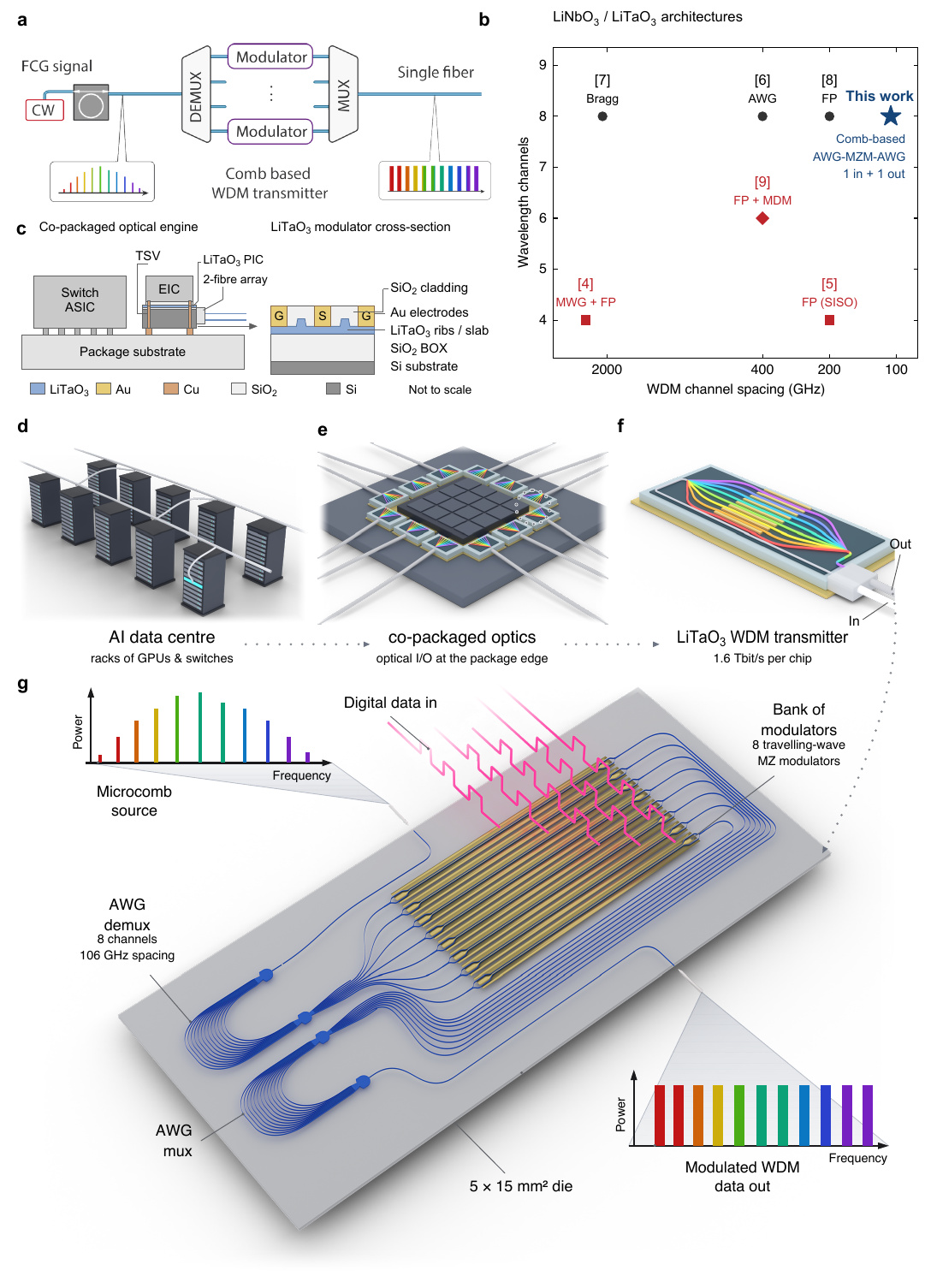}
		\caption{\textbf{Monolithic wavelength-division-multiplexed (WDM) transmitter on thin-film lithium tantalate.} AWG: arrayed waveguide grating. MZM: Mach--Zehnder modulator.
		\textbf{(a)} Comb demultiplexing, independent modulation and recombination into one output.
		\textbf{(b)} Wavelength-channel count and spacing of \LiNbO{} (red) and \LiTaO{} (grey) transmitters. Squares: Liu \emph{et al.}'s multimode waveguide grating (MWG) with Fabry--P\'erot (FP) modulators~\cite{liu_compact_wdm_2023}, and single-input, single-output (SISO) FP~\cite{liu_ln_dwdm_2024}. Circles: Huang \emph{et al.}'s AWG~\cite{huang_toward_2025}, Zhu \emph{et al.}'s Bragg-grating~\cite{zhu_8_2025} and FP~\cite{zhu_8240_2025}. Diamond: Zhu \emph{et al.}'s hybrid FP/mode-division multiplexing (MDM)~\cite{zhu_hybrid_ln_2026} (six wavelengths, six modes each). Blue star: this work, comb-based AWG--MZM--AWG with one input and output. Spacing decreases logarithmically to the right.
		\textbf{(c)} Prospective co-packaged optical engine with copper through-silicon vias (TSVs) and a \LiTaO{} modulator cross-section (not to scale).
		\textbf{(d--f)} AI data-centre interconnects (d), switch-package optical engines (e) and the two-fibre \LiTaO{} transmitter (f).
		\textbf{(g)} The $5\times15$~mm$^2$ die integrates two eight-channel AWGs and eight travelling-wave MZMs on a nominal 106~GHz grid.
		}
		\label{fig:concept}
	\end{paperfigure}

Bringing the optical interface into the package places tighter constraints on the optical engine's area, thermal budget and fibre connections. Wavelength-division multiplexing (WDM) addresses the fibre-count constraint by allowing multiple independently modulated carriers to share each fibre~\cite{buscaino_copackaged_2019,kim_co-packaged_2026}. Scaling the channel count, however, also requires a scalable optical source. In conventional WDM transmitters, arrays of individually stabilized continuous-wave lasers increase power consumption, footprint, cost and control complexity as channels are added. A Kerr-soliton microcomb supplies a phase-coherent grid of carriers from one pump laser and one microresonator, removing the need for a dedicated laser per channel~\cite{delhaye_optical_2007,kippenberg_microresonator-based_2011,brasch_photonic_2016,kippenberg_dissipative_2018,shen_integrated_2020,xiang_laser_2021}. Such sources have enabled links with terabit- and petabit-per-second aggregate transmission rates~\cite{pfeifle_coherent_2014,marin-palomo_microresonator-based_2017,corcoran_ultra-dense_2020,jorgensen_petabit-per-second_2022}.

Using a shared multi-wavelength source in a compact transmitter requires low-loss wavelength separation, independent broadband modulation and on-chip recombination into one output waveguide. The material platform must therefore support both passive wavelength routing and efficient electro-optic (EO) modulation. Silicon photonics offers dense, foundry-scale integration~\cite{li_silicon_2015,shekhar_roadmapping_2024} and has enabled comb-driven transmission~\cite{shu_microcomb-driven_2022,rizzo_massively_2023}. Its conventional plasma-dispersion modulators, however, face a trade-off between modulation efficiency and optical loss~\cite{shekhar_roadmapping_2024}. Lithium niobate (\LiNbO{}) on insulator provides a native Pockels response for efficient broadband EO modulation~\cite{wang_integrated_2018,he_high-performance_2019,boes_lithium_2023,hu_integrated_2025}, although the in-plane anisotropy of X-cut \LiNbO{} complicates wavelength routing~\cite{yi_anisotropy-free_2024}. Integrated \LiNbO{} transmitters have demonstrated four-channel coarse and dense WDM, including a resonant single-input, single-output device on a 200~GHz grid~\cite{chen_cwdm_2022,liu_compact_wdm_2023,liu_ln_dwdm_2024}. Further \LiNbO{} implementations include athermal passive-optical-network routing and mode/wavelength multiplexing~\cite{he_athermal_pon_2025,zhao_modedivisionmultiplexing_2025,zhu_hybrid_ln_2026}.

Lithium tantalate (\LiTaO{}) on insulator combines a Pockels coefficient comparable to \LiNbO{} ($r_{33}\approx30$~pm/V) with roughly an order of magnitude lower birefringence. This lower birefringence suppresses polarization-mode hybridization and facilitates passive photonic integration~\cite{huang_toward_2025}. Arrayed waveguide gratings (AWGs)~\cite{hulyal_arrayed_2025} and broadband Mach--Zehnder modulators (MZMs)~\cite{wang_ultrabroadband_2024,lin_copper_2026} provide the routing and modulation building blocks for \LiTaO{} transmitters. Reported \LiTaO{} WDM devices combine modulation with a single wavelength-routing stage, leaving separate channel ports on one side of the circuit~\cite{zhu_8_2025,zhu_8240_2025,huang_toward_2025}. Integrating the complementary routing stage on chip provides a common optical input and output as the channel count increases. The comparison in \fref{fig:concept}b summarizes the wavelength-channel counts and spacings of selected \LiNbO{} and \LiTaO{} transmitters alongside this work.

Here we demonstrate a monolithic \LiTaO{}-on-insulator WDM transmitter with eight channels on a nominal 106~GHz grid and a single optical input and output. Laser-fed measurements establish terabit-scale operation across all eight channels, while a shared \SiN Kerr-soliton microcomb yields an estimated net data rate of 582.7~Gbit~s$^{-1}$ across four measured channels. The architecture retains a two-fibre interface as the channel count increases, addressing the fibre-count constraint of co-packaged optical engines.


	\section*{Device concept and design}
	The transmitter implements carrier demultiplexing, independent modulation and recombination on chip (\fref{fig:concept}a), providing the optical functions between a shared multi-wavelength source and a WDM transmission fibre. This arrangement replaces separate fibre-coupled routing components and packaged modulators with a compact optical engine suited to the package edge (\fref{fig:concept}d--f)~\cite{tan_co-packaged_2023,kim_co-packaged_2026}. The single-input, single-output architecture provides the photonic building block for the prospective co-packaged engine in \fref{fig:concept}c.

	The circuit is fabricated using the wafer-scale process described in Ref.~\cite{hulyal_arrayed_2025} (Methods). The fabricated wafer, die, waveguide cross-section and transmitter layout are shown in \fref{fig:passive}a--d. On the $5\times15$~mm$^2$ \LiTaO{}-on-insulator die (\fref{fig:concept}g), an eight-channel AWG demultiplexer~\cite{smit_phasar-based_1996} separates the incoming carriers, eight travelling-wave MZMs encode independent data streams, and a second AWG recombines the modulated carriers into one output waveguide. Both AWGs use the validated Rowland star-coupler design (\fref{fig:passive}e), with eight channels on a nominal 106~GHz grid in the telecommunication C band (Supplementary Section~1). The low in-plane anisotropy of \LiTaO{} allows the arrayed waveguides to follow the Rowland geometry without anisotropy compensation. By comparison, X-cut \LiNbO{} requires layouts that compensate for the direction-dependent refractive index~\cite{yi_anisotropy-free_2024}. The two AWGs have identical layouts and are placed immediately adjacent on the die. Film-thickness variations then shift their centre frequencies nearly equally, helping the passbands track one another across the wavelength grid. The eight travelling-wave MZMs are arranged at a 300~\um{} pitch, with independent ground pads to suppress electrical crosstalk between neighbouring channels (\fref{fig:passive}f). The electrode design is described in Supplementary Section~3. Inverse tapers provide fibre coupling (\fref{fig:passive}g), and euler waveguide bends connect the circuit components (\fref{fig:passive}h).

	\begin{paperfigure}
		\centering
		\paperfiguregraphic[height=0.70\textheight,keepaspectratio]{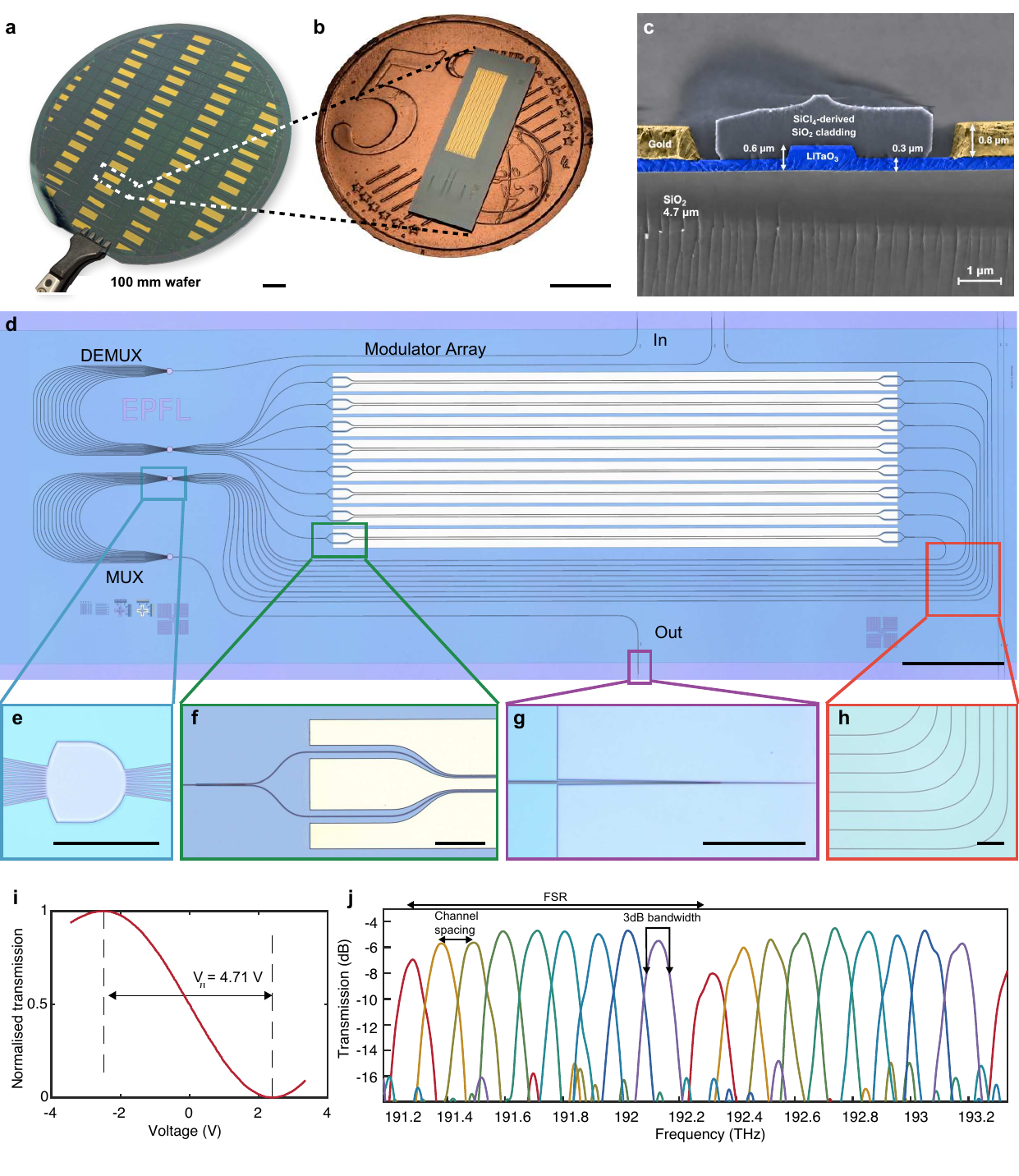}
		\caption{\textbf{Fabricated device and passive and electro-optic characterization.}
		\textbf{(a,b)} The 100-mm \LiTaO{}-on-insulator wafer (a) and a singulated $5\times15$~mm$^2$ die on a one-cent coin (b). Scale bars, 1~cm and 5~mm, respectively.
		\textbf{(c)} False-colour scanning electron micrograph of the \LiTaO{} ridge waveguide, gold electrodes, \ce{SiCl4}-derived \ce{SiO2} cladding and buried \ce{SiO2}. Scale bar, 1~$\mu$m.
		\textbf{(d)} Transmitter micrograph of device D210\_F6\_C1, with traced waveguide paths linking the input, AWG demultiplexer, modulator array, AWG multiplexer and output. Scale bar, 1~mm.
		\textbf{(e--h)} Enlarged views of an AWG star coupler (e), modulator electrodes (f), inverse taper (g) and routing bends (h). Scale bars, 100~$\mu$m.
		\textbf{(i)} Normalized DC modulator transmission, with half-wave voltage $V_\pi=4.71$~V.
		\textbf{(j)} Eight-channel AWG transmission across two free spectral ranges (FSRs), showing a mean channel spacing of 106.2~GHz and a mean 3\,dB bandwidth of 84.4~GHz.
		}
		\label{fig:passive}
	\end{paperfigure}

	\section*{Passive and electro-optic characterization}
	We first characterize the wavelength-routing and modulation functions separately. The measured eight-channel AWG transmission (\fref{fig:passive}j) gives a mean channel spacing of 106.2~GHz, a 3\,dB bandwidth of 84.4~GHz, an insertion loss of $5.29$~dB and adjacent-channel crosstalk of $-13.95$~dB, consistent with the wafer-scale statistics of Ref.~\cite{hulyal_arrayed_2025}. A passive swept-carrier measurement through the complete transmitter yields end-to-end passbands with a mean 3\,dB width of 55.3~GHz ($\sigma=15.7$~GHz, Supplementary Section~2). The integrated modulators have a measured DC half-wave voltage of $V_\pi=4.71$~V (\fref{fig:passive}i), corresponding to $V_\pi L=3.58$~V\,cm.

	The narrower end-to-end passbands arise because the swept, unmodulated carrier passes through both AWGs, so the measured transmission follows the product of their responses. For identical Gaussian power passbands, this cascade reduces the 3\,dB width from 84.4~GHz to $84.4/\sqrt{2}\approx59.7$~GHz, close to the measured 55.3~GHz. During data transmission, the input AWG receives narrow, unmodulated carriers, and the MZMs generate modulation sidebands. The output AWG then filters the modulated signal according to its individual passband, with a mean 3\,dB width of 84.4~GHz (\fref{fig:passive}j).

	\section*{High-speed WDM transmission}
	We first establish high-speed WDM operation using continuous-wave carriers from external-cavity lasers. Two adjacent channels carry independent data sequences simultaneously, including the effect of the driven neighbour (Supplementary Section~7). The results in \fref{fig:transmission} use four disjoint adjacent-pair configurations that cover all eight channels.

	For each pair, the laser carriers are combined and launched into the chip with aligned polarization, and the MZMs are biased at quadrature (\fref{fig:transmission}a). The recombined output is amplified and optically demultiplexed for direct detection, with each channel acquired sequentially. The electrical drive, receiver and digital signal processing are detailed in Supplementary Section~4.

	\begin{paperfigure}
		\centering
		\paperfiguregraphic{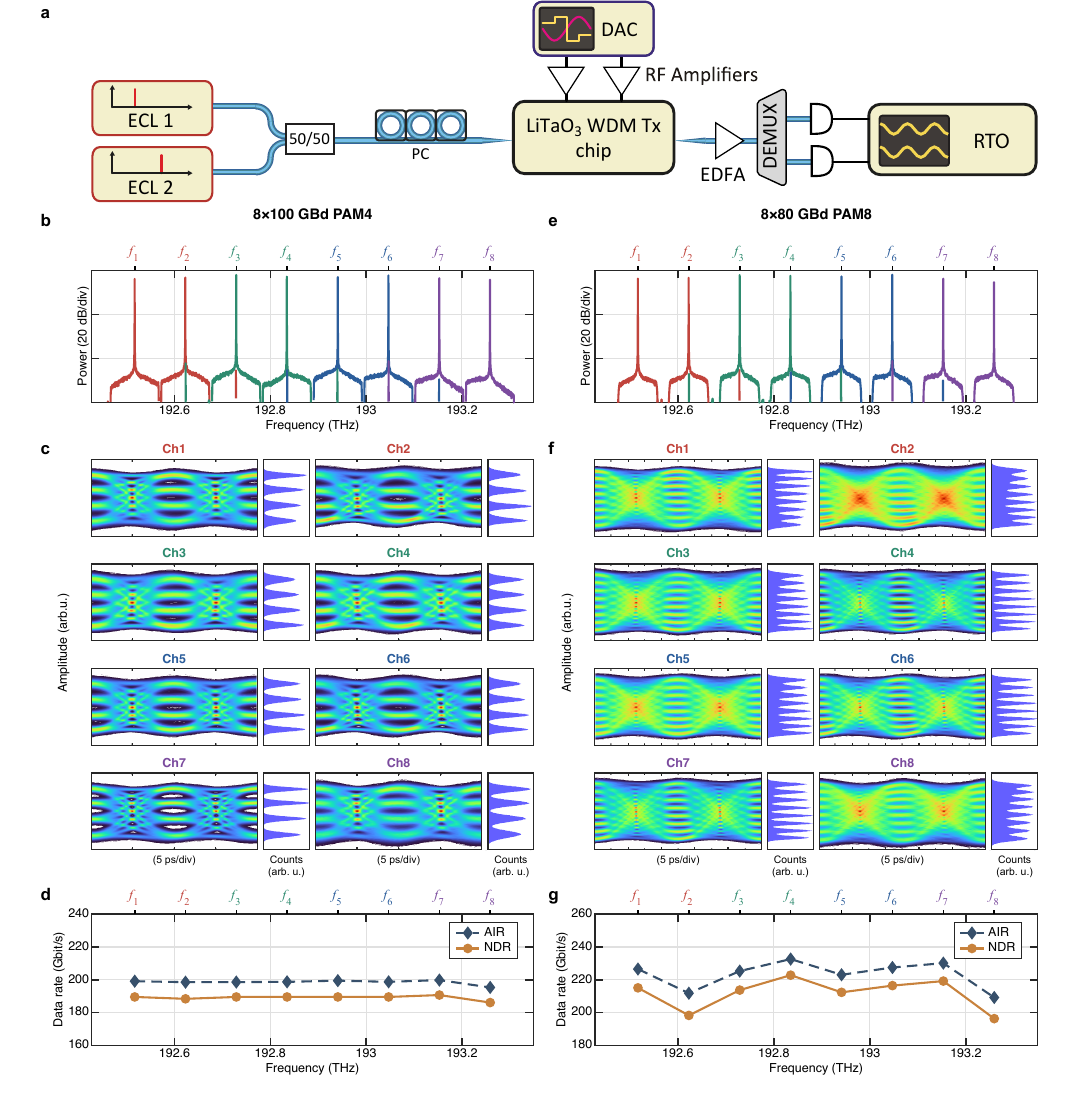}
		\caption{\textbf{Terabit-scale laser-fed WDM transmission.}
		\textbf{(a)} Experimental setup using two external-cavity lasers (ECL), a polarization controller (PC), a digital-to-analogue converter (DAC), RF amplifiers, an erbium-doped fibre amplifier (EDFA), a demultiplexer (DEMUX) and a real-time oscilloscope (RTO). Adjacent channels carry independent data and are measured in four pairs.
		\textbf{(b--d)} Optical spectra (b), recovered eye diagrams and amplitude-count histograms (c), and per-channel achievable information rates (AIR, dashed) and estimated net data rates (NDR, solid) for $100$~GBd PAM4, giving an aggregate line rate of 1.6~Tbit~s$^{-1}$.
		\textbf{(e--g)} Corresponding spectra (e), eyes and histograms (f), and rates (g) for $80$~GBd PAM8, with an aggregate estimated NDR of 1.69~Tbit~s$^{-1}$. Colours identify adjacent channel pairs. Eyes are shown at 5~ps/div.
		}
		\label{fig:transmission}
	\end{paperfigure}

	Using this setup, we generate and receive 100~GBd PAM4 and 80~GBd PAM8 signals across the eight channels~\cite{hulyal_16_2026}. The optical spectra in \fref{fig:transmission}b and e combine the four adjacent-pair measurements, with colours identifying the pairs. The recovered eye diagrams for all eight channels are shown in \fref{fig:transmission}c and f.

	We quantify the recovered signal quality using the channel signal-to-noise ratio (CSNR), generalized mutual information (GMI) and bit-error ratio (BER) before forward-error correction (FEC). The achievable information rate (AIR) is calculated from the GMI and symbol rate, while the estimated net data rate (NDR) accounts for the corresponding FEC code rate (Supplementary Section~4). AIRs (dashed lines) and estimated NDRs (solid lines) are plotted in \fref{fig:transmission}d and g, respectively. All channels have pre-FEC BERs below the reference criterion of $4\times10^{-2}$. Summing the per-channel rates across these four adjacent pairs gives an aggregate PAM4 line rate of 1.6~Tbit~s$^{-1}$ and an estimated PAM8 NDR of 1.6943~Tbit~s$^{-1}$. These results establish high-capacity WDM operation of the integrated transmitter under laser drive.

	\section*{Comb-driven operation}
	Having established operation with laser carriers, we replace them with a dissipative Kerr soliton microcomb~\cite{kippenberg_dissipative_2018}, supplying multiple carriers from a shared source. The comb is generated in a \SiN microresonator on a separate chip, with a nominal 106~GHz free spectral range matched to the AWG grid (\fref{fig:comb}a,b). The single-soliton state remains stable for several hours without thermoelectric or feedback control, and its optical spectrum is shown in \fref{fig:comb}c. After amplification and spectral flattening, eight comb lines aligned to the AWG grid are launched into the transmitter. Four channels are modulated at 80~GBd in adjacent pairs (\fref{fig:comb}d).

	\begin{paperfigure}
		\centering
		\paperfiguregraphic{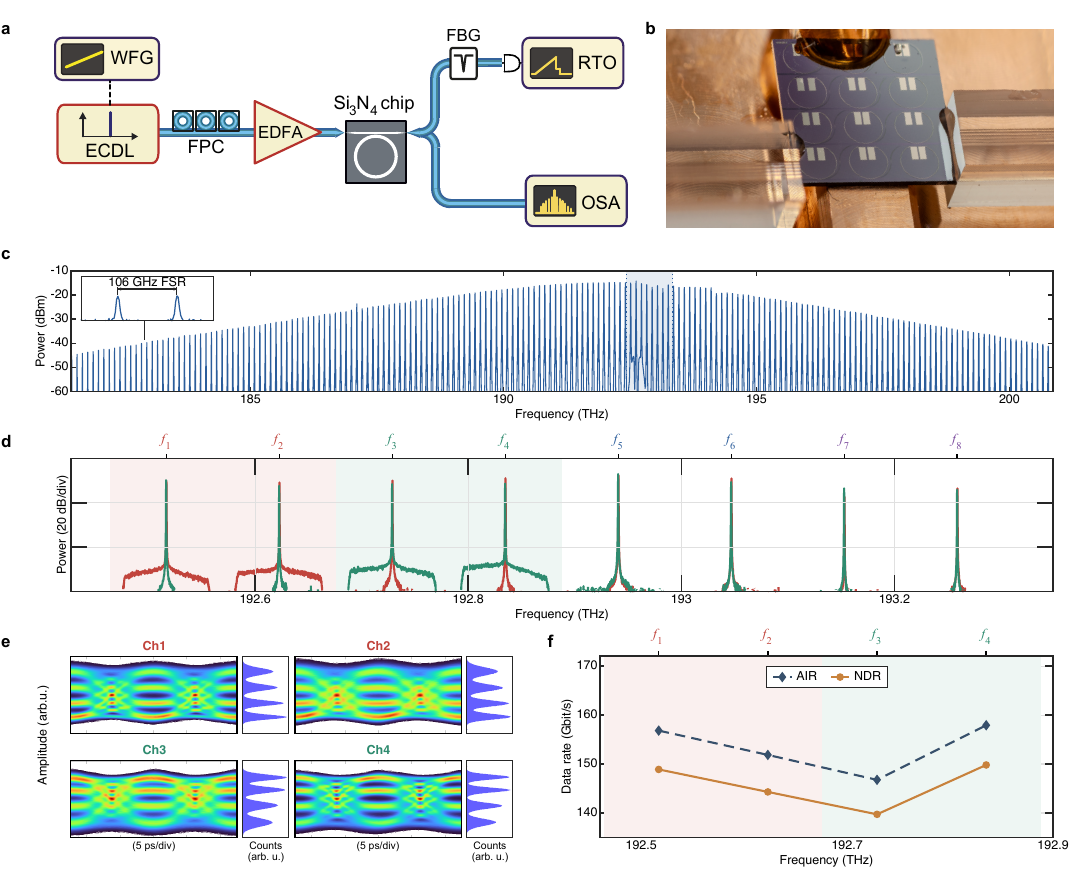}
		\caption{\textbf{Soliton-microcomb-driven WDM transmission.}
		\textbf{(a)} Comb-generation setup: an external-cavity diode laser (ECDL), swept by a waveform generator (WFG), pumps a \SiN microresonator through a fibre polarization controller (FPC) and EDFA. A fibre Bragg grating (FBG) suppresses the pump before detection on the RTO. An optical spectrum analyser (OSA) records the spectrum.
		\textbf{(b)} Fibre-array-packaged \SiN comb chip.
		\textbf{(c)} Single-soliton optical spectrum. The shaded band contains the carriers in (d). The inset shows the 106~GHz comb-line spacing.
		\textbf{(d)} Output spectra for the two modulated pairs, $f_1,f_2$ (terracotta) and $f_3,f_4$ (teal), at 80~GBd PAM4. Channels $f_5$--$f_8$ remain unmodulated. Shading marks the measured channels.
		\textbf{(e)} Recovered eyes at 5~ps/div and amplitude histograms for $f_1$--$f_4$, each histogram normalized to its maximum.
		\textbf{(f)} Per-channel AIR (dashed) and estimated NDR (solid), with an aggregate NDR of 582.7~Gbit~s$^{-1}$.
		}
		\label{fig:comb}
	\end{paperfigure}

	All four characterized channels show open eye diagrams (\fref{fig:comb}e). Their estimated NDRs range from 139.7 to 149.8~Gbit~s$^{-1}$ and sum to 582.7~Gbit~s$^{-1}$ (\fref{fig:comb}f). The GMI values range from 1.83 to 1.97~bit/symbol, corresponding to PAM4-equivalent information rates, with CSNR values of 10--12~dB.

	The source spectra and carrier-power characterization are provided in Supplementary Sections~5 and 6.

	\section*{Discussion and outlook}
	The laser-fed and comb-driven results demonstrate dense WDM transmission through the same monolithic \LiTaO{} circuit. The common optical input accommodates a shared multi-wavelength source, while the integrated routing stages recombine the independently modulated carriers into one output fibre.

	With a nominal 106~GHz spacing, the transmitter provides the densest wavelength grid among the eight-channel \LiTaO{} devices compared here, together with a two-fibre interface (\fref{fig:concept}b). The three prior implementations use 16.8~nm Bragg-filter spacing~\cite{zhu_8_2025}, 200~GHz O-band FP filters~\cite{zhu_8240_2025}, or a 400~GHz AWG multiplexer~\cite{huang_toward_2025}, each with eight separate channel ports and one common port. Correspondingly, the broadband travelling-wave MZMs in our transmitter provide high-speed modulation without per-channel modulator resonance locking.

	In the prospective co-packaged implementation shown in \fref{fig:concept}c, electronic drivers would be stacked above the transmitter to shorten the high-frequency electrical connections to its MZMs~\cite{kim_co-packaged_2026}. The common optical input and output would allow each WDM engine to retain a two-fibre interface as its wavelength-channel count increases, limiting the number of fibre attachments at the package edge. Copper damascene processing has already been demonstrated for \LiTaO{} modulators, providing planarized metallization suitable for chip-on-wafer integration with driver electronics~\cite{lin_copper_2026}. Adapting this process to the present AWG--MZM--AWG circuit would provide a manufacturing route towards this co-packaged engine.

	The monolithic architecture can extend to higher channel counts through larger AWGs while retaining its two-fibre interface. Beyond the eight-channel demonstration, a reticle-scale array could target around 100 modulators in a $20\times20$~mm$^2$ field. At the demonstrated 300~\um{} pitch and $7.6$~mm electrode length, two banks of 50 modulators would occupy approximately $15\times15.2$~mm$^2$ in electrode-array area. This geometrical estimate provides a starting point for multiple WDM engines, each with a common optical input and output. A complete reticle layout must also accommodate wavelength routing, electrical contacts and optical interfaces. For example, extending the demonstrated $200$~Gbit~s$^{-1}$ gross line rate per channel to 50 active channels distributed across multiple WDM engines on a single reticle would yield $10$~Tbit~s$^{-1}$ aggregate throughput. Retaining a common optical input and output for each WDM engine reduces the number of fibre connections compared with designs that couple individual channels separately, supporting high shoreline bandwidth density for co-packaged optical interconnects. The \LiTaO{} platform also draws on wafer supply chains established for RF filters~\cite{wang_lithium_2024}, supporting the manufacturing scale required for co-packaged optical interconnects in AI clusters~\cite{shekhar_roadmapping_2024,kim_co-packaged_2026}.

	\section*{Methods}

	\par\noindent\textbf{Device fabrication.} \quad
		The transmitter was fabricated using the wafer-scale process described in Ref.~\cite{hulyal_arrayed_2025}. The starting wafer (D210) comprised a 600~nm X-cut \LiTaO{} film on 4.7~\um{} of buried \SiO and a 700~\um{} silicon substrate. Deep ultraviolet (DUV) lithography defined the waveguides in a diamond-like-carbon hardmask, and ion-beam etching transferred the pattern into the \LiTaO{} film, leaving a $\sim$300~nm slab. A second DUV lithography step defined the double-layer inverse-taper couplers. Subsequent wet etching in an alkaline solution smoothed the waveguide sidewalls, after which dense \SiO cladding was deposited by inductively coupled plasma chemical vapour deposition using silicon tetrachloride (\ce{SiCl4}) as the silicon precursor. The gold travelling-wave electrodes were patterned by DUV lithography and lift-off. After removal of the residual photoresist, the chip boundaries were defined, followed by dry etching with fluorine-based chemistry and backside wafer grinding to release the chips. The resulting die was edge-coupled through the inverse tapers~\cite{hulyal_arrayed_2025}.

	\medskip
	\begin{footnotesize}

		\noindent \textbf{Author Contributions}: S.U.H. conceived the idea, designed and simulated the transmitter circuit, and performed the passive characterization. J.C. fabricated the \LiTaO{} transmitter chips. X.J. fabricated the \SiN microresonator chip used for soliton microcomb generation. J.S. packaged the \SiN microcomb device. D.F. performed the high-speed and comb-driven transmission experiments and the digital signal processing, with support from A.K. S.U.H. and D.F. analysed the data. S.U.H. wrote the manuscript with input from all authors. C.K. and T.J.K. supervised the project. S.U.H., D.F. and J.C. contributed equally to this work.

		\noindent \textbf{Funding Information and Disclaimer}: This work was supported by Contract N660012424006 (NaPSAC) from the Defense Advanced Research Projects Agency (DARPA), Defense Sciences Office (DSO).  It was further supported by the Swiss National Science Foundation under grant agreement No. 21649 (HEROIC).  S.U.H. acknowledges support from Horizon Europe research and innovation programme under the Marie Skłodowska-Curie grant agreement No. 101119968 (MicrocombSys) and from the Swiss State Secretariat for Education, Research and Innovation (SERI).

		\noindent \textbf{Acknowledgments}:
		The samples were fabricated in the EPFL Center of MicroNanoTechnology (CMi). We thank Junyin Zhang for helping with the microwave characterization measurements.

		\noindent \textbf{Competing Interests}: T.J.K. is a co-founder and shareholder of Luxtelligence SA, a foundry commercializing \LiNbO{} and \LiTaO{} photonic integrated circuits (PICs), as well as DEEPLIGHT SA, a start-up commercializing PIC-based frequency-agile low-noise lasers. The other authors declare no competing interests.

		\noindent \textbf{Data Availability Statement}: The code and data used to produce the plots in this work will be made available on \texttt{Zenodo} upon publication.

		\noindent\textbf{Correspondence and requests for materials} should be addressed to T.J.K.
	\end{footnotesize}
	\bibliography{main_refs}

\begin{thebibliography}{38}%
\makeatletter
\providecommand \@ifxundefined [1]{%
 \@ifx{#1\undefined}
}%
\providecommand \@ifnum [1]{%
 \ifnum #1\expandafter \@firstoftwo
 \else \expandafter \@secondoftwo
 \fi
}%
\providecommand \@ifx [1]{%
 \ifx #1\expandafter \@firstoftwo
 \else \expandafter \@secondoftwo
 \fi
}%
\providecommand \natexlab [1]{#1}%
\providecommand \enquote  [1]{``#1''}%
\providecommand \bibnamefont  [1]{#1}%
\providecommand \bibfnamefont [1]{#1}%
\providecommand \citenamefont [1]{#1}%
\providecommand \href@noop [0]{\@secondoftwo}%
\providecommand \href [0]{\begingroup \@sanitize@url \@href}%
\providecommand \@href[1]{\@@startlink{#1}\@@href}%
\providecommand \@@href[1]{\endgroup#1\@@endlink}%
\providecommand \@sanitize@url [0]{\catcode `\\12\catcode `\$12\catcode
  `\&12\catcode `\#12\catcode `\^12\catcode `\_12\catcode `\%12\relax}%
\providecommand \@@startlink[1]{}%
\providecommand \@@endlink[0]{}%
\providecommand \url  [0]{\begingroup\@sanitize@url \@url }%
\providecommand \@url [1]{\endgroup\@href {#1}{\urlprefix }}%
\providecommand \urlprefix  [0]{URL }%
\providecommand \Eprint [0]{\href }%
\providecommand \doibase [0]{https://doi.org/}%
\providecommand \selectlanguage [0]{\@gobble}%
\providecommand \bibinfo  [0]{\@secondoftwo}%
\providecommand \bibfield  [0]{\@secondoftwo}%
\providecommand \translation [1]{[#1]}%
\providecommand \BibitemOpen [0]{}%
\providecommand \bibitemStop [0]{}%
\providecommand \bibitemNoStop [0]{.\EOS\space}%
\providecommand \EOS [0]{\spacefactor3000\relax}%
\providecommand \BibitemShut  [1]{\csname bibitem#1\endcsname}%
\let\auto@bib@innerbib\@empty
\bibitem [{\citenamefont {Kim}\ \emph {et~al.}(2026)\citenamefont {Kim},
  \citenamefont {Choi}, \citenamefont {Zograf}, \citenamefont {Yoo},
  \citenamefont {Baek}, \citenamefont {Kim}, \citenamefont {Kim}, \citenamefont
  {Kim}, \citenamefont {Chae}, \citenamefont {Kim}, \citenamefont {Hong},\ and\
  \citenamefont {Lee}}]{kim_co-packaged_2026}%
  \BibitemOpen
  \bibfield  {author} {\bibinfo {author} {\bibfnamefont {B.}~\bibnamefont
  {Kim}}, \bibinfo {author} {\bibfnamefont {S.~H.}\ \bibnamefont {Choi}},
  \bibinfo {author} {\bibfnamefont {G.}~\bibnamefont {Zograf}}, \bibinfo
  {author} {\bibfnamefont {Y.~J.}\ \bibnamefont {Yoo}}, \bibinfo {author}
  {\bibfnamefont {Y.}~\bibnamefont {Baek}}, \bibinfo {author} {\bibfnamefont
  {S.}~\bibnamefont {Kim}}, \bibinfo {author} {\bibfnamefont {J.}~\bibnamefont
  {Kim}}, \bibinfo {author} {\bibfnamefont {J.}~\bibnamefont {Kim}}, \bibinfo
  {author} {\bibfnamefont {S.~H.}\ \bibnamefont {Chae}}, \bibinfo {author}
  {\bibfnamefont {H.}~\bibnamefont {Kim}}, \bibinfo {author} {\bibfnamefont
  {S.}~\bibnamefont {Hong}},\ and\ \bibinfo {author} {\bibfnamefont
  {K.}~\bibnamefont {Lee}},\ }\href
  {https://doi.org/10.1038/s41928-026-01681-6} {\bibfield  {journal} {\bibinfo
  {journal} {Nature Electronics}\ }\textbf {\bibinfo {volume} {9}},\ \bibinfo
  {pages} {853} (\bibinfo {year} {2026})}\BibitemShut {NoStop}%
\bibitem [{\citenamefont {Tan}\ \emph {et~al.}(2023)\citenamefont {Tan},
  \citenamefont {Xu}, \citenamefont {Liu}, \citenamefont {Feng}, \citenamefont
  {Zhang}, \citenamefont {Yao}, \citenamefont {Chen}, \citenamefont {Guo},
  \citenamefont {Han}, \citenamefont {Wen}, \citenamefont {Chen}, \citenamefont
  {He}, \citenamefont {Zheng}, \citenamefont {Ming}, \citenamefont {Tu},
  \citenamefont {Fu}, \citenamefont {Qi}, \citenamefont {Li}, \citenamefont
  {Geng}, \citenamefont {Wen}, \citenamefont {Yang}, \citenamefont {He},
  \citenamefont {Liu}, \citenamefont {Xue}, \citenamefont {Wang}, \citenamefont
  {Qiu}, \citenamefont {Mi}, \citenamefont {Li}, \citenamefont {Chang},
  \citenamefont {Lai}, \citenamefont {Zhang}, \citenamefont {Hao},\ and\
  \citenamefont {Qin}}]{tan_co-packaged_2023}%
  \BibitemOpen
  \bibfield  {author} {\bibinfo {author} {\bibfnamefont {M.}~\bibnamefont
  {Tan}}, \bibinfo {author} {\bibfnamefont {J.}~\bibnamefont {Xu}}, \bibinfo
  {author} {\bibfnamefont {S.}~\bibnamefont {Liu}}, \bibinfo {author}
  {\bibfnamefont {J.}~\bibnamefont {Feng}}, \bibinfo {author} {\bibfnamefont
  {H.}~\bibnamefont {Zhang}}, \bibinfo {author} {\bibfnamefont
  {C.}~\bibnamefont {Yao}}, \bibinfo {author} {\bibfnamefont {S.}~\bibnamefont
  {Chen}}, \bibinfo {author} {\bibfnamefont {H.}~\bibnamefont {Guo}}, \bibinfo
  {author} {\bibfnamefont {G.}~\bibnamefont {Han}}, \bibinfo {author}
  {\bibfnamefont {Z.}~\bibnamefont {Wen}}, \bibinfo {author} {\bibfnamefont
  {B.}~\bibnamefont {Chen}}, \bibinfo {author} {\bibfnamefont {Y.}~\bibnamefont
  {He}}, \bibinfo {author} {\bibfnamefont {X.}~\bibnamefont {Zheng}}, \bibinfo
  {author} {\bibfnamefont {D.}~\bibnamefont {Ming}}, \bibinfo {author}
  {\bibfnamefont {Y.}~\bibnamefont {Tu}}, \bibinfo {author} {\bibfnamefont
  {Q.}~\bibnamefont {Fu}}, \bibinfo {author} {\bibfnamefont {N.}~\bibnamefont
  {Qi}}, \bibinfo {author} {\bibfnamefont {D.}~\bibnamefont {Li}}, \bibinfo
  {author} {\bibfnamefont {L.}~\bibnamefont {Geng}}, \bibinfo {author}
  {\bibfnamefont {S.}~\bibnamefont {Wen}}, \bibinfo {author} {\bibfnamefont
  {F.}~\bibnamefont {Yang}}, \bibinfo {author} {\bibfnamefont {H.}~\bibnamefont
  {He}}, \bibinfo {author} {\bibfnamefont {F.}~\bibnamefont {Liu}}, \bibinfo
  {author} {\bibfnamefont {H.}~\bibnamefont {Xue}}, \bibinfo {author}
  {\bibfnamefont {Y.}~\bibnamefont {Wang}}, \bibinfo {author} {\bibfnamefont
  {C.}~\bibnamefont {Qiu}}, \bibinfo {author} {\bibfnamefont {G.}~\bibnamefont
  {Mi}}, \bibinfo {author} {\bibfnamefont {Y.}~\bibnamefont {Li}}, \bibinfo
  {author} {\bibfnamefont {T.}~\bibnamefont {Chang}}, \bibinfo {author}
  {\bibfnamefont {M.}~\bibnamefont {Lai}}, \bibinfo {author} {\bibfnamefont
  {L.}~\bibnamefont {Zhang}}, \bibinfo {author} {\bibfnamefont
  {Q.}~\bibnamefont {Hao}},\ and\ \bibinfo {author} {\bibfnamefont
  {M.}~\bibnamefont {Qin}},\ }\href
  {https://doi.org/10.1007/s12200-022-00055-y} {\bibfield  {journal} {\bibinfo
  {journal} {Frontiers of Optoelectronics}\ }\textbf {\bibinfo {volume} {16}},\
  \bibinfo {pages} {1} (\bibinfo {year} {2023})}\BibitemShut {NoStop}%
\bibitem [{\citenamefont {Miller}(2017)}]{miller_attojoule_2017}%
  \BibitemOpen
  \bibfield  {author} {\bibinfo {author} {\bibfnamefont {D.~A.~B.}\
  \bibnamefont {Miller}},\ }\href {https://doi.org/10.1109/JLT.2017.2647779}
  {\bibfield  {journal} {\bibinfo  {journal} {Journal of Lightwave Technology}\
  }\textbf {\bibinfo {volume} {35}},\ \bibinfo {pages} {346} (\bibinfo {year}
  {2017})}\BibitemShut {NoStop}%
\bibitem [{\citenamefont {Liu}\ \emph {et~al.}(2023)\citenamefont {Liu},
  \citenamefont {Pan}, \citenamefont {Huang}, \citenamefont {He}, \citenamefont
  {Zhang}, \citenamefont {Yu}, \citenamefont {Liu}, \citenamefont {Shi},\ and\
  \citenamefont {Dai}}]{liu_compact_wdm_2023}%
  \BibitemOpen
  \bibfield  {author} {\bibinfo {author} {\bibfnamefont {H.}~\bibnamefont
  {Liu}}, \bibinfo {author} {\bibfnamefont {B.}~\bibnamefont {Pan}}, \bibinfo
  {author} {\bibfnamefont {Y.}~\bibnamefont {Huang}}, \bibinfo {author}
  {\bibfnamefont {J.}~\bibnamefont {He}}, \bibinfo {author} {\bibfnamefont
  {M.}~\bibnamefont {Zhang}}, \bibinfo {author} {\bibfnamefont
  {Z.}~\bibnamefont {Yu}}, \bibinfo {author} {\bibfnamefont {L.}~\bibnamefont
  {Liu}}, \bibinfo {author} {\bibfnamefont {Y.}~\bibnamefont {Shi}},\ and\
  \bibinfo {author} {\bibfnamefont {D.}~\bibnamefont {Dai}},\ }\href
  {https://doi.org/10.37188/lam.2023.013} {\bibfield  {journal} {\bibinfo
  {journal} {Light: Advanced Manufacturing}\ }\textbf {\bibinfo {volume} {4}},\
  \bibinfo {pages} {133} (\bibinfo {year} {2023})}\BibitemShut {NoStop}%
\bibitem [{\citenamefont {Liu}\ \emph {et~al.}(2024)\citenamefont {Liu},
  \citenamefont {Pan}, \citenamefont {Li}, \citenamefont {Yu}, \citenamefont
  {Liu}, \citenamefont {Shi},\ and\ \citenamefont {Dai}}]{liu_ln_dwdm_2024}%
  \BibitemOpen
  \bibfield  {author} {\bibinfo {author} {\bibfnamefont {H.}~\bibnamefont
  {Liu}}, \bibinfo {author} {\bibfnamefont {B.}~\bibnamefont {Pan}}, \bibinfo
  {author} {\bibfnamefont {H.}~\bibnamefont {Li}}, \bibinfo {author}
  {\bibfnamefont {Z.}~\bibnamefont {Yu}}, \bibinfo {author} {\bibfnamefont
  {L.}~\bibnamefont {Liu}}, \bibinfo {author} {\bibfnamefont {Y.}~\bibnamefont
  {Shi}},\ and\ \bibinfo {author} {\bibfnamefont {D.}~\bibnamefont {Dai}},\
  }\href {https://doi.org/10.1117/1.AP.6.6.066001} {\bibfield  {journal}
  {\bibinfo  {journal} {Advanced Photonics}\ }\textbf {\bibinfo {volume} {6}},\
  \bibinfo {pages} {066001} (\bibinfo {year} {2024})}\BibitemShut {NoStop}%
\bibitem [{\citenamefont {Huang}\ \emph {et~al.}(2025)\citenamefont {Huang},
  \citenamefont {Shen}, \citenamefont {Wang}, \citenamefont {Xu}, \citenamefont
  {Liu}, \citenamefont {Wang}, \citenamefont {Gao}, \citenamefont {Yao},
  \citenamefont {Cao}, \citenamefont {Chen}, \citenamefont {Wang},
  \citenamefont {Zhang}, \citenamefont {Wu}, \citenamefont {Zhu}, \citenamefont
  {Xiong}, \citenamefont {Zhao}, \citenamefont {Li}, \citenamefont {Yu},
  \citenamefont {Liu}, \citenamefont {Shi},\ and\ \citenamefont
  {Dai}}]{huang_toward_2025}%
  \BibitemOpen
  \bibfield  {author} {\bibinfo {author} {\bibfnamefont {F.}~\bibnamefont
  {Huang}}, \bibinfo {author} {\bibfnamefont {X.}~\bibnamefont {Shen}},
  \bibinfo {author} {\bibfnamefont {S.}~\bibnamefont {Wang}}, \bibinfo {author}
  {\bibfnamefont {H.}~\bibnamefont {Xu}}, \bibinfo {author} {\bibfnamefont
  {H.}~\bibnamefont {Liu}}, \bibinfo {author} {\bibfnamefont {Z.}~\bibnamefont
  {Wang}}, \bibinfo {author} {\bibfnamefont {H.}~\bibnamefont {Gao}}, \bibinfo
  {author} {\bibfnamefont {X.}~\bibnamefont {Yao}}, \bibinfo {author}
  {\bibfnamefont {H.}~\bibnamefont {Cao}}, \bibinfo {author} {\bibfnamefont
  {B.}~\bibnamefont {Chen}}, \bibinfo {author} {\bibfnamefont {X.}~\bibnamefont
  {Wang}}, \bibinfo {author} {\bibfnamefont {J.}~\bibnamefont {Zhang}},
  \bibinfo {author} {\bibfnamefont {Z.}~\bibnamefont {Wu}}, \bibinfo {author}
  {\bibfnamefont {M.}~\bibnamefont {Zhu}}, \bibinfo {author} {\bibfnamefont
  {H.}~\bibnamefont {Xiong}}, \bibinfo {author} {\bibfnamefont
  {W.}~\bibnamefont {Zhao}}, \bibinfo {author} {\bibfnamefont {H.}~\bibnamefont
  {Li}}, \bibinfo {author} {\bibfnamefont {Z.}~\bibnamefont {Yu}}, \bibinfo
  {author} {\bibfnamefont {L.}~\bibnamefont {Liu}}, \bibinfo {author}
  {\bibfnamefont {Y.}~\bibnamefont {Shi}},\ and\ \bibinfo {author}
  {\bibfnamefont {D.}~\bibnamefont {Dai}},\ }\href
  {https://doi.org/10.1002/advs.202410345} {\bibfield  {journal} {\bibinfo
  {journal} {Advanced Science}\ }\textbf {\bibinfo {volume} {12}},\ \bibinfo
  {pages} {2410345} (\bibinfo {year} {2025})}\BibitemShut {NoStop}%
\bibitem [{\citenamefont {Zhu}\ \emph {et~al.}(2025{\natexlab{a}})\citenamefont
  {Zhu}, \citenamefont {Huang}, \citenamefont {Liu}, \citenamefont {Wang},
  \citenamefont {Gao}, \citenamefont {Zhao}, \citenamefont {Zhao},
  \citenamefont {Lian}, \citenamefont {Gao}, \citenamefont {Yu},\ and\
  \citenamefont {Dai}}]{zhu_8_2025}%
  \BibitemOpen
  \bibfield  {author} {\bibinfo {author} {\bibfnamefont {M.}~\bibnamefont
  {Zhu}}, \bibinfo {author} {\bibfnamefont {F.}~\bibnamefont {Huang}}, \bibinfo
  {author} {\bibfnamefont {D.}~\bibnamefont {Liu}}, \bibinfo {author}
  {\bibfnamefont {W.}~\bibnamefont {Wang}}, \bibinfo {author} {\bibfnamefont
  {A.}~\bibnamefont {Gao}}, \bibinfo {author} {\bibfnamefont {W.}~\bibnamefont
  {Zhao}}, \bibinfo {author} {\bibfnamefont {S.}~\bibnamefont {Zhao}}, \bibinfo
  {author} {\bibfnamefont {D.}~\bibnamefont {Lian}}, \bibinfo {author}
  {\bibfnamefont {C.}~\bibnamefont {Gao}}, \bibinfo {author} {\bibfnamefont
  {Z.}~\bibnamefont {Yu}},\ and\ \bibinfo {author} {\bibfnamefont
  {D.}~\bibnamefont {Dai}},\ }\href
  {https://doi.org/10.1186/s43074-025-00183-6} {\bibfield  {journal} {\bibinfo
  {journal} {PhotoniX}\ }\textbf {\bibinfo {volume} {6}},\ \bibinfo {pages}
  {27} (\bibinfo {year} {2025}{\natexlab{a}})}\BibitemShut {NoStop}%
\bibitem [{\citenamefont {Zhu}\ \emph {et~al.}(2025{\natexlab{b}})\citenamefont
  {Zhu}, \citenamefont {Liu}, \citenamefont {Wang}, \citenamefont {Zhong},
  \citenamefont {Huang}, \citenamefont {Zhao}, \citenamefont {Gao},
  \citenamefont {Gan}, \citenamefont {Zhao}, \citenamefont {Gao}, \citenamefont
  {Lian}, \citenamefont {Li}, \citenamefont {Yu},\ and\ \citenamefont
  {Dai}}]{zhu_8240_2025}%
  \BibitemOpen
  \bibfield  {author} {\bibinfo {author} {\bibfnamefont {M.}~\bibnamefont
  {Zhu}}, \bibinfo {author} {\bibfnamefont {D.}~\bibnamefont {Liu}}, \bibinfo
  {author} {\bibfnamefont {W.}~\bibnamefont {Wang}}, \bibinfo {author}
  {\bibfnamefont {H.}~\bibnamefont {Zhong}}, \bibinfo {author} {\bibfnamefont
  {F.}~\bibnamefont {Huang}}, \bibinfo {author} {\bibfnamefont
  {S.}~\bibnamefont {Zhao}}, \bibinfo {author} {\bibfnamefont {A.}~\bibnamefont
  {Gao}}, \bibinfo {author} {\bibfnamefont {J.}~\bibnamefont {Gan}}, \bibinfo
  {author} {\bibfnamefont {W.}~\bibnamefont {Zhao}}, \bibinfo {author}
  {\bibfnamefont {C.}~\bibnamefont {Gao}}, \bibinfo {author} {\bibfnamefont
  {D.}~\bibnamefont {Lian}}, \bibinfo {author} {\bibfnamefont {H.}~\bibnamefont
  {Li}}, \bibinfo {author} {\bibfnamefont {Z.}~\bibnamefont {Yu}},\ and\
  \bibinfo {author} {\bibfnamefont {D.}~\bibnamefont {Dai}},\ }\href
  {https://doi.org/10.1038/s41467-025-66370-y} {\bibfield  {journal} {\bibinfo
  {journal} {Nature Communications}\ }\textbf {\bibinfo {volume} {16}},\
  \bibinfo {pages} {11365} (\bibinfo {year} {2025}{\natexlab{b}})}\BibitemShut
  {NoStop}%
\bibitem [{\citenamefont {Zhu}\ \emph {et~al.}(2026)\citenamefont {Zhu},
  \citenamefont {Wang}, \citenamefont {Ma}, \citenamefont {Gao}, \citenamefont
  {Gao}, \citenamefont {Wang}, \citenamefont {Huang}, \citenamefont {Bao},
  \citenamefont {Liu}, \citenamefont {Gan}, \citenamefont {Yu}, \citenamefont
  {Li}, \citenamefont {Zhao},\ and\ \citenamefont {Dai}}]{zhu_hybrid_ln_2026}%
  \BibitemOpen
  \bibfield  {author} {\bibinfo {author} {\bibfnamefont {M.}~\bibnamefont
  {Zhu}}, \bibinfo {author} {\bibfnamefont {W.}~\bibnamefont {Wang}}, \bibinfo
  {author} {\bibfnamefont {R.}~\bibnamefont {Ma}}, \bibinfo {author}
  {\bibfnamefont {A.}~\bibnamefont {Gao}}, \bibinfo {author} {\bibfnamefont
  {C.}~\bibnamefont {Gao}}, \bibinfo {author} {\bibfnamefont {Z.}~\bibnamefont
  {Wang}}, \bibinfo {author} {\bibfnamefont {F.}~\bibnamefont {Huang}},
  \bibinfo {author} {\bibfnamefont {Z.}~\bibnamefont {Bao}}, \bibinfo {author}
  {\bibfnamefont {D.}~\bibnamefont {Liu}}, \bibinfo {author} {\bibfnamefont
  {J.}~\bibnamefont {Gan}}, \bibinfo {author} {\bibfnamefont {Z.}~\bibnamefont
  {Yu}}, \bibinfo {author} {\bibfnamefont {H.}~\bibnamefont {Li}}, \bibinfo
  {author} {\bibfnamefont {W.}~\bibnamefont {Zhao}},\ and\ \bibinfo {author}
  {\bibfnamefont {D.}~\bibnamefont {Dai}},\ }\href
  {https://doi.org/10.29026/oea.2026.250341} {\bibfield  {journal} {\bibinfo
  {journal} {Opto-Electronic Advances}\ }\textbf {\bibinfo {volume} {9}},\
  \bibinfo {pages} {250341} (\bibinfo {year} {2026})}\BibitemShut {NoStop}%
\bibitem [{\citenamefont {Buscaino}\ \emph {et~al.}(2019)\citenamefont
  {Buscaino}, \citenamefont {Taylor},\ and\ \citenamefont
  {Kahn}}]{buscaino_copackaged_2019}%
  \BibitemOpen
  \bibfield  {author} {\bibinfo {author} {\bibfnamefont {B.}~\bibnamefont
  {Buscaino}}, \bibinfo {author} {\bibfnamefont {B.~D.}\ \bibnamefont
  {Taylor}},\ and\ \bibinfo {author} {\bibfnamefont {J.~M.}\ \bibnamefont
  {Kahn}},\ }\href {https://doi.org/10.1109/JLT.2019.2916988} {\bibfield
  {journal} {\bibinfo  {journal} {Journal of Lightwave Technology}\ }\textbf
  {\bibinfo {volume} {37}},\ \bibinfo {pages} {3401} (\bibinfo {year}
  {2019})}\BibitemShut {NoStop}%
\bibitem [{\citenamefont {Del'Haye}\ \emph {et~al.}(2007)\citenamefont
  {Del'Haye}, \citenamefont {Schliesser}, \citenamefont {Arcizet},
  \citenamefont {Wilken}, \citenamefont {Holzwarth},\ and\ \citenamefont
  {Kippenberg}}]{delhaye_optical_2007}%
  \BibitemOpen
  \bibfield  {author} {\bibinfo {author} {\bibfnamefont {P.}~\bibnamefont
  {Del'Haye}}, \bibinfo {author} {\bibfnamefont {A.}~\bibnamefont
  {Schliesser}}, \bibinfo {author} {\bibfnamefont {O.}~\bibnamefont {Arcizet}},
  \bibinfo {author} {\bibfnamefont {T.}~\bibnamefont {Wilken}}, \bibinfo
  {author} {\bibfnamefont {R.}~\bibnamefont {Holzwarth}},\ and\ \bibinfo
  {author} {\bibfnamefont {T.~J.}\ \bibnamefont {Kippenberg}},\ }\href
  {https://doi.org/10.1038/nature06401} {\bibfield  {journal} {\bibinfo
  {journal} {Nature}\ }\textbf {\bibinfo {volume} {450}},\ \bibinfo {pages}
  {1214} (\bibinfo {year} {2007})}\BibitemShut {NoStop}%
\bibitem [{\citenamefont {Kippenberg}\ \emph {et~al.}(2011)\citenamefont
  {Kippenberg}, \citenamefont {Holzwarth},\ and\ \citenamefont
  {Diddams}}]{kippenberg_microresonator-based_2011}%
  \BibitemOpen
  \bibfield  {author} {\bibinfo {author} {\bibfnamefont {T.~J.}\ \bibnamefont
  {Kippenberg}}, \bibinfo {author} {\bibfnamefont {R.}~\bibnamefont
  {Holzwarth}},\ and\ \bibinfo {author} {\bibfnamefont {S.~A.}\ \bibnamefont
  {Diddams}},\ }\href {https://doi.org/10.1126/science.1193968} {\bibfield
  {journal} {\bibinfo  {journal} {Science}\ }\textbf {\bibinfo {volume}
  {332}},\ \bibinfo {pages} {555} (\bibinfo {year} {2011})}\BibitemShut
  {NoStop}%
\bibitem [{\citenamefont {Brasch}\ \emph {et~al.}(2016)\citenamefont {Brasch},
  \citenamefont {Geiselmann}, \citenamefont {Herr}, \citenamefont {Lihachev},
  \citenamefont {Pfeiffer}, \citenamefont {Gorodetsky},\ and\ \citenamefont
  {Kippenberg}}]{brasch_photonic_2016}%
  \BibitemOpen
  \bibfield  {author} {\bibinfo {author} {\bibfnamefont {V.}~\bibnamefont
  {Brasch}}, \bibinfo {author} {\bibfnamefont {M.}~\bibnamefont {Geiselmann}},
  \bibinfo {author} {\bibfnamefont {T.}~\bibnamefont {Herr}}, \bibinfo {author}
  {\bibfnamefont {G.}~\bibnamefont {Lihachev}}, \bibinfo {author}
  {\bibfnamefont {M.~H.~P.}\ \bibnamefont {Pfeiffer}}, \bibinfo {author}
  {\bibfnamefont {M.~L.}\ \bibnamefont {Gorodetsky}},\ and\ \bibinfo {author}
  {\bibfnamefont {T.~J.}\ \bibnamefont {Kippenberg}},\ }\href
  {https://doi.org/10.1126/science.aad4811} {\bibfield  {journal} {\bibinfo
  {journal} {Science}\ }\textbf {\bibinfo {volume} {351}},\ \bibinfo {pages}
  {357} (\bibinfo {year} {2016})}\BibitemShut {NoStop}%
\bibitem [{\citenamefont {Kippenberg}\ \emph {et~al.}(2018)\citenamefont
  {Kippenberg}, \citenamefont {Gaeta}, \citenamefont {Lipson},\ and\
  \citenamefont {Gorodetsky}}]{kippenberg_dissipative_2018}%
  \BibitemOpen
  \bibfield  {author} {\bibinfo {author} {\bibfnamefont {T.~J.}\ \bibnamefont
  {Kippenberg}}, \bibinfo {author} {\bibfnamefont {A.~L.}\ \bibnamefont
  {Gaeta}}, \bibinfo {author} {\bibfnamefont {M.}~\bibnamefont {Lipson}},\ and\
  \bibinfo {author} {\bibfnamefont {M.~L.}\ \bibnamefont {Gorodetsky}},\ }\href
  {https://doi.org/10.1126/science.aan8083} {\bibfield  {journal} {\bibinfo
  {journal} {Science}\ }\textbf {\bibinfo {volume} {361}},\ \bibinfo {pages}
  {eaan8083} (\bibinfo {year} {2018})}\BibitemShut {NoStop}%
\bibitem [{\citenamefont {Shen}\ \emph {et~al.}(2020)\citenamefont {Shen},
  \citenamefont {Chang}, \citenamefont {Liu}, \citenamefont {Wang},
  \citenamefont {Yang}, \citenamefont {Xiang}, \citenamefont {Wang},
  \citenamefont {He}, \citenamefont {Liu}, \citenamefont {Xie}, \citenamefont
  {Guo}, \citenamefont {Kinghorn}, \citenamefont {Wu}, \citenamefont {Ji},
  \citenamefont {Kippenberg}, \citenamefont {Vahala},\ and\ \citenamefont
  {Bowers}}]{shen_integrated_2020}%
  \BibitemOpen
  \bibfield  {author} {\bibinfo {author} {\bibfnamefont {B.}~\bibnamefont
  {Shen}}, \bibinfo {author} {\bibfnamefont {L.}~\bibnamefont {Chang}},
  \bibinfo {author} {\bibfnamefont {J.}~\bibnamefont {Liu}}, \bibinfo {author}
  {\bibfnamefont {H.}~\bibnamefont {Wang}}, \bibinfo {author} {\bibfnamefont
  {Q.-F.}\ \bibnamefont {Yang}}, \bibinfo {author} {\bibfnamefont
  {C.}~\bibnamefont {Xiang}}, \bibinfo {author} {\bibfnamefont {R.~N.}\
  \bibnamefont {Wang}}, \bibinfo {author} {\bibfnamefont {J.}~\bibnamefont
  {He}}, \bibinfo {author} {\bibfnamefont {T.}~\bibnamefont {Liu}}, \bibinfo
  {author} {\bibfnamefont {W.}~\bibnamefont {Xie}}, \bibinfo {author}
  {\bibfnamefont {J.}~\bibnamefont {Guo}}, \bibinfo {author} {\bibfnamefont
  {D.}~\bibnamefont {Kinghorn}}, \bibinfo {author} {\bibfnamefont
  {L.}~\bibnamefont {Wu}}, \bibinfo {author} {\bibfnamefont {Q.-X.}\
  \bibnamefont {Ji}}, \bibinfo {author} {\bibfnamefont {T.~J.}\ \bibnamefont
  {Kippenberg}}, \bibinfo {author} {\bibfnamefont {K.}~\bibnamefont {Vahala}},\
  and\ \bibinfo {author} {\bibfnamefont {J.~E.}\ \bibnamefont {Bowers}},\
  }\href {https://doi.org/10.1038/s41586-020-2358-x} {\bibfield  {journal}
  {\bibinfo  {journal} {Nature}\ }\textbf {\bibinfo {volume} {582}},\ \bibinfo
  {pages} {365} (\bibinfo {year} {2020})}\BibitemShut {NoStop}%
\bibitem [{\citenamefont {Xiang}\ \emph {et~al.}(2021)\citenamefont {Xiang},
  \citenamefont {Liu}, \citenamefont {Guo}, \citenamefont {Chang},
  \citenamefont {Wang}, \citenamefont {Weng}, \citenamefont {Peters},
  \citenamefont {Xie}, \citenamefont {Zhang}, \citenamefont {Riemensberger},
  \citenamefont {Selvidge}, \citenamefont {Kippenberg},\ and\ \citenamefont
  {Bowers}}]{xiang_laser_2021}%
  \BibitemOpen
  \bibfield  {author} {\bibinfo {author} {\bibfnamefont {C.}~\bibnamefont
  {Xiang}}, \bibinfo {author} {\bibfnamefont {J.}~\bibnamefont {Liu}}, \bibinfo
  {author} {\bibfnamefont {J.}~\bibnamefont {Guo}}, \bibinfo {author}
  {\bibfnamefont {L.}~\bibnamefont {Chang}}, \bibinfo {author} {\bibfnamefont
  {R.~N.}\ \bibnamefont {Wang}}, \bibinfo {author} {\bibfnamefont
  {W.}~\bibnamefont {Weng}}, \bibinfo {author} {\bibfnamefont {J.}~\bibnamefont
  {Peters}}, \bibinfo {author} {\bibfnamefont {W.}~\bibnamefont {Xie}},
  \bibinfo {author} {\bibfnamefont {Z.}~\bibnamefont {Zhang}}, \bibinfo
  {author} {\bibfnamefont {J.}~\bibnamefont {Riemensberger}}, \bibinfo {author}
  {\bibfnamefont {J.}~\bibnamefont {Selvidge}}, \bibinfo {author}
  {\bibfnamefont {T.~J.}\ \bibnamefont {Kippenberg}},\ and\ \bibinfo {author}
  {\bibfnamefont {J.~E.}\ \bibnamefont {Bowers}},\ }\href
  {https://doi.org/10.1126/science.abh2076} {\bibfield  {journal} {\bibinfo
  {journal} {Science}\ }\textbf {\bibinfo {volume} {373}},\ \bibinfo {pages}
  {99} (\bibinfo {year} {2021})}\BibitemShut {NoStop}%
\bibitem [{\citenamefont {Pfeifle}\ \emph {et~al.}(2014)\citenamefont
  {Pfeifle}, \citenamefont {Brasch}, \citenamefont {Lauermann}, \citenamefont
  {Yu}, \citenamefont {Wegner}, \citenamefont {Herr}, \citenamefont
  {Hartinger}, \citenamefont {Schindler}, \citenamefont {Li}, \citenamefont
  {Hillerkuss}, \citenamefont {Schmogrow}, \citenamefont {Weimann},
  \citenamefont {Holzwarth}, \citenamefont {Freude}, \citenamefont {Leuthold},
  \citenamefont {Kippenberg},\ and\ \citenamefont
  {Koos}}]{pfeifle_coherent_2014}%
  \BibitemOpen
  \bibfield  {author} {\bibinfo {author} {\bibfnamefont {J.}~\bibnamefont
  {Pfeifle}}, \bibinfo {author} {\bibfnamefont {V.}~\bibnamefont {Brasch}},
  \bibinfo {author} {\bibfnamefont {M.}~\bibnamefont {Lauermann}}, \bibinfo
  {author} {\bibfnamefont {Y.}~\bibnamefont {Yu}}, \bibinfo {author}
  {\bibfnamefont {D.}~\bibnamefont {Wegner}}, \bibinfo {author} {\bibfnamefont
  {T.}~\bibnamefont {Herr}}, \bibinfo {author} {\bibfnamefont {K.}~\bibnamefont
  {Hartinger}}, \bibinfo {author} {\bibfnamefont {P.}~\bibnamefont
  {Schindler}}, \bibinfo {author} {\bibfnamefont {J.}~\bibnamefont {Li}},
  \bibinfo {author} {\bibfnamefont {D.}~\bibnamefont {Hillerkuss}}, \bibinfo
  {author} {\bibfnamefont {R.}~\bibnamefont {Schmogrow}}, \bibinfo {author}
  {\bibfnamefont {C.}~\bibnamefont {Weimann}}, \bibinfo {author} {\bibfnamefont
  {R.}~\bibnamefont {Holzwarth}}, \bibinfo {author} {\bibfnamefont
  {W.}~\bibnamefont {Freude}}, \bibinfo {author} {\bibfnamefont
  {J.}~\bibnamefont {Leuthold}}, \bibinfo {author} {\bibfnamefont {T.~J.}\
  \bibnamefont {Kippenberg}},\ and\ \bibinfo {author} {\bibfnamefont
  {C.}~\bibnamefont {Koos}},\ }\href {https://doi.org/10.1038/nphoton.2014.57}
  {\bibfield  {journal} {\bibinfo  {journal} {Nature Photonics}\ }\textbf
  {\bibinfo {volume} {8}},\ \bibinfo {pages} {375} (\bibinfo {year}
  {2014})}\BibitemShut {NoStop}%
\bibitem [{\citenamefont {Marin-Palomo}\ \emph {et~al.}(2017)\citenamefont
  {Marin-Palomo}, \citenamefont {Kemal}, \citenamefont {Karpov}, \citenamefont
  {Kordts}, \citenamefont {Pfeifle}, \citenamefont {Pfeiffer}, \citenamefont
  {Trocha}, \citenamefont {Wolf}, \citenamefont {Brasch}, \citenamefont
  {Anderson}, \citenamefont {Rosenberger}, \citenamefont {Vijayan},
  \citenamefont {Freude}, \citenamefont {Kippenberg},\ and\ \citenamefont
  {Koos}}]{marin-palomo_microresonator-based_2017}%
  \BibitemOpen
  \bibfield  {author} {\bibinfo {author} {\bibfnamefont {P.}~\bibnamefont
  {Marin-Palomo}}, \bibinfo {author} {\bibfnamefont {J.~N.}\ \bibnamefont
  {Kemal}}, \bibinfo {author} {\bibfnamefont {M.}~\bibnamefont {Karpov}},
  \bibinfo {author} {\bibfnamefont {A.}~\bibnamefont {Kordts}}, \bibinfo
  {author} {\bibfnamefont {J.}~\bibnamefont {Pfeifle}}, \bibinfo {author}
  {\bibfnamefont {M.~H.~P.}\ \bibnamefont {Pfeiffer}}, \bibinfo {author}
  {\bibfnamefont {P.}~\bibnamefont {Trocha}}, \bibinfo {author} {\bibfnamefont
  {S.}~\bibnamefont {Wolf}}, \bibinfo {author} {\bibfnamefont {V.}~\bibnamefont
  {Brasch}}, \bibinfo {author} {\bibfnamefont {M.~H.}\ \bibnamefont
  {Anderson}}, \bibinfo {author} {\bibfnamefont {R.}~\bibnamefont
  {Rosenberger}}, \bibinfo {author} {\bibfnamefont {K.}~\bibnamefont
  {Vijayan}}, \bibinfo {author} {\bibfnamefont {W.}~\bibnamefont {Freude}},
  \bibinfo {author} {\bibfnamefont {T.~J.}\ \bibnamefont {Kippenberg}},\ and\
  \bibinfo {author} {\bibfnamefont {C.}~\bibnamefont {Koos}},\ }\href
  {https://doi.org/10.1038/nature22387} {\bibfield  {journal} {\bibinfo
  {journal} {Nature}\ }\textbf {\bibinfo {volume} {546}},\ \bibinfo {pages}
  {274} (\bibinfo {year} {2017})}\BibitemShut {NoStop}%
\bibitem [{\citenamefont {Corcoran}\ \emph {et~al.}(2020)\citenamefont
  {Corcoran}, \citenamefont {Tan}, \citenamefont {Xu}, \citenamefont {Boes},
  \citenamefont {Wu}, \citenamefont {Nguyen}, \citenamefont {Chu},
  \citenamefont {Little}, \citenamefont {Morandotti}, \citenamefont
  {Mitchell},\ and\ \citenamefont {Moss}}]{corcoran_ultra-dense_2020}%
  \BibitemOpen
  \bibfield  {author} {\bibinfo {author} {\bibfnamefont {B.}~\bibnamefont
  {Corcoran}}, \bibinfo {author} {\bibfnamefont {M.}~\bibnamefont {Tan}},
  \bibinfo {author} {\bibfnamefont {X.}~\bibnamefont {Xu}}, \bibinfo {author}
  {\bibfnamefont {A.}~\bibnamefont {Boes}}, \bibinfo {author} {\bibfnamefont
  {J.}~\bibnamefont {Wu}}, \bibinfo {author} {\bibfnamefont {T.~G.}\
  \bibnamefont {Nguyen}}, \bibinfo {author} {\bibfnamefont {S.~T.}\
  \bibnamefont {Chu}}, \bibinfo {author} {\bibfnamefont {B.~E.}\ \bibnamefont
  {Little}}, \bibinfo {author} {\bibfnamefont {R.}~\bibnamefont {Morandotti}},
  \bibinfo {author} {\bibfnamefont {A.}~\bibnamefont {Mitchell}},\ and\
  \bibinfo {author} {\bibfnamefont {D.~J.}\ \bibnamefont {Moss}},\ }\href
  {https://doi.org/10.1038/s41467-020-16265-x} {\bibfield  {journal} {\bibinfo
  {journal} {Nature Communications}\ }\textbf {\bibinfo {volume} {11}},\
  \bibinfo {pages} {2568} (\bibinfo {year} {2020})}\BibitemShut {NoStop}%
\bibitem [{\citenamefont {J{\o}rgensen}\ \emph {et~al.}(2022)\citenamefont
  {J{\o}rgensen}, \citenamefont {Kong}, \citenamefont {Henriksen},
  \citenamefont {Klejs}, \citenamefont {Ye}, \citenamefont {Helgason},
  \citenamefont {Hansen}, \citenamefont {Hu}, \citenamefont {Yankov},
  \citenamefont {Forchhammer}, \citenamefont {Andrekson}, \citenamefont
  {Larsson}, \citenamefont {Karlsson}, \citenamefont {Schr{\"{o}}der},
  \citenamefont {Sasaki}, \citenamefont {Aikawa}, \citenamefont {Thomsen},
  \citenamefont {Morioka}, \citenamefont {Galili}, \citenamefont
  {Torres-Company},\ and\ \citenamefont
  {Oxenl{\o}we}}]{jorgensen_petabit-per-second_2022}%
  \BibitemOpen
  \bibfield  {author} {\bibinfo {author} {\bibfnamefont {A.~A.}\ \bibnamefont
  {J{\o}rgensen}}, \bibinfo {author} {\bibfnamefont {D.}~\bibnamefont {Kong}},
  \bibinfo {author} {\bibfnamefont {M.~R.}\ \bibnamefont {Henriksen}}, \bibinfo
  {author} {\bibfnamefont {F.}~\bibnamefont {Klejs}}, \bibinfo {author}
  {\bibfnamefont {Z.}~\bibnamefont {Ye}}, \bibinfo {author} {\bibfnamefont
  {{\'{O}}.~B.}\ \bibnamefont {Helgason}}, \bibinfo {author} {\bibfnamefont
  {H.~E.}\ \bibnamefont {Hansen}}, \bibinfo {author} {\bibfnamefont
  {H.}~\bibnamefont {Hu}}, \bibinfo {author} {\bibfnamefont {M.}~\bibnamefont
  {Yankov}}, \bibinfo {author} {\bibfnamefont {S.}~\bibnamefont {Forchhammer}},
  \bibinfo {author} {\bibfnamefont {P.}~\bibnamefont {Andrekson}}, \bibinfo
  {author} {\bibfnamefont {A.}~\bibnamefont {Larsson}}, \bibinfo {author}
  {\bibfnamefont {M.}~\bibnamefont {Karlsson}}, \bibinfo {author}
  {\bibfnamefont {J.}~\bibnamefont {Schr{\"{o}}der}}, \bibinfo {author}
  {\bibfnamefont {Y.}~\bibnamefont {Sasaki}}, \bibinfo {author} {\bibfnamefont
  {K.}~\bibnamefont {Aikawa}}, \bibinfo {author} {\bibfnamefont {J.~W.}\
  \bibnamefont {Thomsen}}, \bibinfo {author} {\bibfnamefont {T.}~\bibnamefont
  {Morioka}}, \bibinfo {author} {\bibfnamefont {M.}~\bibnamefont {Galili}},
  \bibinfo {author} {\bibfnamefont {V.}~\bibnamefont {Torres-Company}},\ and\
  \bibinfo {author} {\bibfnamefont {L.~K.}\ \bibnamefont {Oxenl{\o}we}},\
  }\href {https://doi.org/10.1038/s41566-022-01082-z} {\bibfield  {journal}
  {\bibinfo  {journal} {Nature Photonics}\ }\textbf {\bibinfo {volume} {16}},\
  \bibinfo {pages} {798} (\bibinfo {year} {2022})}\BibitemShut {NoStop}%
\bibitem [{\citenamefont {Li}\ \emph {et~al.}(2015)\citenamefont {Li},
  \citenamefont {Zhang}, \citenamefont {Zhang},\ and\ \citenamefont
  {Poon}}]{li_silicon_2015}%
  \BibitemOpen
  \bibfield  {author} {\bibinfo {author} {\bibfnamefont {Y.}~\bibnamefont
  {Li}}, \bibinfo {author} {\bibfnamefont {Y.}~\bibnamefont {Zhang}}, \bibinfo
  {author} {\bibfnamefont {L.}~\bibnamefont {Zhang}},\ and\ \bibinfo {author}
  {\bibfnamefont {A.~W.}\ \bibnamefont {Poon}},\ }\href
  {https://doi.org/10.1364/PRJ.3.000B10} {\bibfield  {journal} {\bibinfo
  {journal} {Photonics Research}\ }\textbf {\bibinfo {volume} {3}},\ \bibinfo
  {pages} {B10} (\bibinfo {year} {2015})}\BibitemShut {NoStop}%
\bibitem [{\citenamefont {Shekhar}\ \emph {et~al.}(2024)\citenamefont
  {Shekhar}, \citenamefont {Bogaerts}, \citenamefont {Chrostowski},
  \citenamefont {Bowers}, \citenamefont {Hochberg}, \citenamefont {Soref},\
  and\ \citenamefont {Shastri}}]{shekhar_roadmapping_2024}%
  \BibitemOpen
  \bibfield  {author} {\bibinfo {author} {\bibfnamefont {S.}~\bibnamefont
  {Shekhar}}, \bibinfo {author} {\bibfnamefont {W.}~\bibnamefont {Bogaerts}},
  \bibinfo {author} {\bibfnamefont {L.}~\bibnamefont {Chrostowski}}, \bibinfo
  {author} {\bibfnamefont {J.~E.}\ \bibnamefont {Bowers}}, \bibinfo {author}
  {\bibfnamefont {M.}~\bibnamefont {Hochberg}}, \bibinfo {author}
  {\bibfnamefont {R.}~\bibnamefont {Soref}},\ and\ \bibinfo {author}
  {\bibfnamefont {B.~J.}\ \bibnamefont {Shastri}},\ }\href
  {https://doi.org/10.1038/s41467-024-44750-0} {\bibfield  {journal} {\bibinfo
  {journal} {Nature Communications}\ }\textbf {\bibinfo {volume} {15}},\
  \bibinfo {pages} {751} (\bibinfo {year} {2024})}\BibitemShut {NoStop}%
\bibitem [{\citenamefont {Shu}\ \emph {et~al.}(2022)\citenamefont {Shu},
  \citenamefont {Chang}, \citenamefont {Tao}, \citenamefont {Shen},
  \citenamefont {Xie}, \citenamefont {Jin}, \citenamefont {Netherton},
  \citenamefont {Tao}, \citenamefont {Zhang}, \citenamefont {Chen},
  \citenamefont {Bai}, \citenamefont {Qin}, \citenamefont {Yu}, \citenamefont
  {Wang},\ and\ \citenamefont {Bowers}}]{shu_microcomb-driven_2022}%
  \BibitemOpen
  \bibfield  {author} {\bibinfo {author} {\bibfnamefont {H.}~\bibnamefont
  {Shu}}, \bibinfo {author} {\bibfnamefont {L.}~\bibnamefont {Chang}}, \bibinfo
  {author} {\bibfnamefont {Y.}~\bibnamefont {Tao}}, \bibinfo {author}
  {\bibfnamefont {B.}~\bibnamefont {Shen}}, \bibinfo {author} {\bibfnamefont
  {W.}~\bibnamefont {Xie}}, \bibinfo {author} {\bibfnamefont {M.}~\bibnamefont
  {Jin}}, \bibinfo {author} {\bibfnamefont {A.}~\bibnamefont {Netherton}},
  \bibinfo {author} {\bibfnamefont {Z.}~\bibnamefont {Tao}}, \bibinfo {author}
  {\bibfnamefont {X.}~\bibnamefont {Zhang}}, \bibinfo {author} {\bibfnamefont
  {R.}~\bibnamefont {Chen}}, \bibinfo {author} {\bibfnamefont {B.}~\bibnamefont
  {Bai}}, \bibinfo {author} {\bibfnamefont {J.}~\bibnamefont {Qin}}, \bibinfo
  {author} {\bibfnamefont {S.}~\bibnamefont {Yu}}, \bibinfo {author}
  {\bibfnamefont {X.}~\bibnamefont {Wang}},\ and\ \bibinfo {author}
  {\bibfnamefont {J.~E.}\ \bibnamefont {Bowers}},\ }\href
  {https://doi.org/10.1038/s41586-022-04579-3} {\bibfield  {journal} {\bibinfo
  {journal} {Nature}\ }\textbf {\bibinfo {volume} {605}},\ \bibinfo {pages}
  {457} (\bibinfo {year} {2022})}\BibitemShut {NoStop}%
\bibitem [{\citenamefont {Rizzo}\ \emph {et~al.}(2023)\citenamefont {Rizzo},
  \citenamefont {Novick}, \citenamefont {Gopal}, \citenamefont {Kim},
  \citenamefont {Ji}, \citenamefont {Daudlin}, \citenamefont {Okawachi},
  \citenamefont {Cheng}, \citenamefont {Lipson}, \citenamefont {Gaeta},\ and\
  \citenamefont {Bergman}}]{rizzo_massively_2023}%
  \BibitemOpen
  \bibfield  {author} {\bibinfo {author} {\bibfnamefont {A.}~\bibnamefont
  {Rizzo}}, \bibinfo {author} {\bibfnamefont {A.}~\bibnamefont {Novick}},
  \bibinfo {author} {\bibfnamefont {V.}~\bibnamefont {Gopal}}, \bibinfo
  {author} {\bibfnamefont {B.~Y.}\ \bibnamefont {Kim}}, \bibinfo {author}
  {\bibfnamefont {X.}~\bibnamefont {Ji}}, \bibinfo {author} {\bibfnamefont
  {S.}~\bibnamefont {Daudlin}}, \bibinfo {author} {\bibfnamefont
  {Y.}~\bibnamefont {Okawachi}}, \bibinfo {author} {\bibfnamefont
  {Q.}~\bibnamefont {Cheng}}, \bibinfo {author} {\bibfnamefont
  {M.}~\bibnamefont {Lipson}}, \bibinfo {author} {\bibfnamefont {A.~L.}\
  \bibnamefont {Gaeta}},\ and\ \bibinfo {author} {\bibfnamefont
  {K.}~\bibnamefont {Bergman}},\ }\href
  {https://doi.org/10.1038/s41566-023-01244-7} {\bibfield  {journal} {\bibinfo
  {journal} {Nature Photonics}\ }\textbf {\bibinfo {volume} {17}},\ \bibinfo
  {pages} {781} (\bibinfo {year} {2023})}\BibitemShut {NoStop}%
\bibitem [{\citenamefont {Wang}\ \emph {et~al.}(2018)\citenamefont {Wang},
  \citenamefont {Zhang}, \citenamefont {Chen}, \citenamefont {Bertrand},
  \citenamefont {Shams-Ansari}, \citenamefont {Chandrasekhar}, \citenamefont
  {Winzer},\ and\ \citenamefont {Lon{\v{c}}ar}}]{wang_integrated_2018}%
  \BibitemOpen
  \bibfield  {author} {\bibinfo {author} {\bibfnamefont {C.}~\bibnamefont
  {Wang}}, \bibinfo {author} {\bibfnamefont {M.}~\bibnamefont {Zhang}},
  \bibinfo {author} {\bibfnamefont {X.}~\bibnamefont {Chen}}, \bibinfo {author}
  {\bibfnamefont {M.}~\bibnamefont {Bertrand}}, \bibinfo {author}
  {\bibfnamefont {A.}~\bibnamefont {Shams-Ansari}}, \bibinfo {author}
  {\bibfnamefont {S.}~\bibnamefont {Chandrasekhar}}, \bibinfo {author}
  {\bibfnamefont {P.}~\bibnamefont {Winzer}},\ and\ \bibinfo {author}
  {\bibfnamefont {M.}~\bibnamefont {Lon{\v{c}}ar}},\ }\href
  {https://doi.org/10.1038/s41586-018-0551-y} {\bibfield  {journal} {\bibinfo
  {journal} {Nature}\ }\textbf {\bibinfo {volume} {562}},\ \bibinfo {pages}
  {101} (\bibinfo {year} {2018})}\BibitemShut {NoStop}%
\bibitem [{\citenamefont {He}\ \emph {et~al.}(2019)\citenamefont {He},
  \citenamefont {Xu}, \citenamefont {Ren}, \citenamefont {Jian}, \citenamefont
  {Ruan}, \citenamefont {Xu}, \citenamefont {Gao}, \citenamefont {Sun},
  \citenamefont {Wen}, \citenamefont {Zhou}, \citenamefont {Liu}, \citenamefont
  {Guo}, \citenamefont {Chen}, \citenamefont {Yu}, \citenamefont {Liu},\ and\
  \citenamefont {Cai}}]{he_high-performance_2019}%
  \BibitemOpen
  \bibfield  {author} {\bibinfo {author} {\bibfnamefont {M.}~\bibnamefont
  {He}}, \bibinfo {author} {\bibfnamefont {M.}~\bibnamefont {Xu}}, \bibinfo
  {author} {\bibfnamefont {Y.}~\bibnamefont {Ren}}, \bibinfo {author}
  {\bibfnamefont {J.}~\bibnamefont {Jian}}, \bibinfo {author} {\bibfnamefont
  {Z.}~\bibnamefont {Ruan}}, \bibinfo {author} {\bibfnamefont {Y.}~\bibnamefont
  {Xu}}, \bibinfo {author} {\bibfnamefont {S.}~\bibnamefont {Gao}}, \bibinfo
  {author} {\bibfnamefont {S.}~\bibnamefont {Sun}}, \bibinfo {author}
  {\bibfnamefont {X.}~\bibnamefont {Wen}}, \bibinfo {author} {\bibfnamefont
  {L.}~\bibnamefont {Zhou}}, \bibinfo {author} {\bibfnamefont {L.}~\bibnamefont
  {Liu}}, \bibinfo {author} {\bibfnamefont {C.}~\bibnamefont {Guo}}, \bibinfo
  {author} {\bibfnamefont {H.}~\bibnamefont {Chen}}, \bibinfo {author}
  {\bibfnamefont {S.}~\bibnamefont {Yu}}, \bibinfo {author} {\bibfnamefont
  {L.}~\bibnamefont {Liu}},\ and\ \bibinfo {author} {\bibfnamefont
  {X.}~\bibnamefont {Cai}},\ }\href {https://doi.org/10.1038/s41566-019-0378-6}
  {\bibfield  {journal} {\bibinfo  {journal} {Nature Photonics}\ }\textbf
  {\bibinfo {volume} {13}},\ \bibinfo {pages} {359} (\bibinfo {year}
  {2019})}\BibitemShut {NoStop}%
\bibitem [{\citenamefont {Boes}\ \emph {et~al.}(2023)\citenamefont {Boes},
  \citenamefont {Chang}, \citenamefont {Langrock}, \citenamefont {Yu},
  \citenamefont {Zhang}, \citenamefont {Lin}, \citenamefont {Lon{\v{c}}ar},
  \citenamefont {Fejer}, \citenamefont {Bowers},\ and\ \citenamefont
  {Mitchell}}]{boes_lithium_2023}%
  \BibitemOpen
  \bibfield  {author} {\bibinfo {author} {\bibfnamefont {A.}~\bibnamefont
  {Boes}}, \bibinfo {author} {\bibfnamefont {L.}~\bibnamefont {Chang}},
  \bibinfo {author} {\bibfnamefont {C.}~\bibnamefont {Langrock}}, \bibinfo
  {author} {\bibfnamefont {M.}~\bibnamefont {Yu}}, \bibinfo {author}
  {\bibfnamefont {M.}~\bibnamefont {Zhang}}, \bibinfo {author} {\bibfnamefont
  {Q.}~\bibnamefont {Lin}}, \bibinfo {author} {\bibfnamefont {M.}~\bibnamefont
  {Lon{\v{c}}ar}}, \bibinfo {author} {\bibfnamefont {M.}~\bibnamefont {Fejer}},
  \bibinfo {author} {\bibfnamefont {J.}~\bibnamefont {Bowers}},\ and\ \bibinfo
  {author} {\bibfnamefont {A.}~\bibnamefont {Mitchell}},\ }\href
  {https://doi.org/10.1126/science.abj4396} {\bibfield  {journal} {\bibinfo
  {journal} {Science}\ }\textbf {\bibinfo {volume} {379}},\ \bibinfo {pages}
  {eabj4396} (\bibinfo {year} {2023})}\BibitemShut {NoStop}%
\bibitem [{\citenamefont {Hu}\ \emph {et~al.}(2025)\citenamefont {Hu},
  \citenamefont {Zhu}, \citenamefont {Lu}, \citenamefont {Zhu}, \citenamefont
  {Song}, \citenamefont {Renaud}, \citenamefont {Assumpcao}, \citenamefont
  {Cheng}, \citenamefont {Xin}, \citenamefont {Yeh}, \citenamefont {Warner},
  \citenamefont {Guo}, \citenamefont {Shams-Ansari}, \citenamefont {Barton},
  \citenamefont {Sinclair},\ and\ \citenamefont
  {Lon{\v{c}}ar}}]{hu_integrated_2025}%
  \BibitemOpen
  \bibfield  {author} {\bibinfo {author} {\bibfnamefont {Y.}~\bibnamefont
  {Hu}}, \bibinfo {author} {\bibfnamefont {D.}~\bibnamefont {Zhu}}, \bibinfo
  {author} {\bibfnamefont {S.}~\bibnamefont {Lu}}, \bibinfo {author}
  {\bibfnamefont {X.}~\bibnamefont {Zhu}}, \bibinfo {author} {\bibfnamefont
  {Y.}~\bibnamefont {Song}}, \bibinfo {author} {\bibfnamefont {D.}~\bibnamefont
  {Renaud}}, \bibinfo {author} {\bibfnamefont {D.}~\bibnamefont {Assumpcao}},
  \bibinfo {author} {\bibfnamefont {R.}~\bibnamefont {Cheng}}, \bibinfo
  {author} {\bibfnamefont {C.~J.}\ \bibnamefont {Xin}}, \bibinfo {author}
  {\bibfnamefont {M.}~\bibnamefont {Yeh}}, \bibinfo {author} {\bibfnamefont
  {H.}~\bibnamefont {Warner}}, \bibinfo {author} {\bibfnamefont
  {X.}~\bibnamefont {Guo}}, \bibinfo {author} {\bibfnamefont {A.}~\bibnamefont
  {Shams-Ansari}}, \bibinfo {author} {\bibfnamefont {D.}~\bibnamefont
  {Barton}}, \bibinfo {author} {\bibfnamefont {N.}~\bibnamefont {Sinclair}},\
  and\ \bibinfo {author} {\bibfnamefont {M.}~\bibnamefont {Lon{\v{c}}ar}},\
  }\href {https://doi.org/10.1038/s42254-025-00825-5} {\bibfield  {journal}
  {\bibinfo  {journal} {Nature Reviews Physics}\ }\textbf {\bibinfo {volume}
  {7}},\ \bibinfo {pages} {237} (\bibinfo {year} {2025})},\ \bibinfo {note}
  {publisher correction:
  \href{https://doi.org/10.1038/s42254-025-00858-w}{10.1038/s42254-025-00858-w}}\BibitemShut
  {NoStop}%
\bibitem [{\citenamefont {Yi}\ \emph {et~al.}(2024)\citenamefont {Yi},
  \citenamefont {Guo}, \citenamefont {Ruan}, \citenamefont {Chen},
  \citenamefont {Wei}, \citenamefont {Lu}, \citenamefont {Gong}, \citenamefont
  {Pan}, \citenamefont {Shen}, \citenamefont {Guan}, \citenamefont {Dai},
  \citenamefont {Zhong},\ and\ \citenamefont {Liu}}]{yi_anisotropy-free_2024}%
  \BibitemOpen
  \bibfield  {author} {\bibinfo {author} {\bibfnamefont {J.}~\bibnamefont
  {Yi}}, \bibinfo {author} {\bibfnamefont {C.}~\bibnamefont {Guo}}, \bibinfo
  {author} {\bibfnamefont {Z.}~\bibnamefont {Ruan}}, \bibinfo {author}
  {\bibfnamefont {G.}~\bibnamefont {Chen}}, \bibinfo {author} {\bibfnamefont
  {H.}~\bibnamefont {Wei}}, \bibinfo {author} {\bibfnamefont {L.}~\bibnamefont
  {Lu}}, \bibinfo {author} {\bibfnamefont {S.}~\bibnamefont {Gong}}, \bibinfo
  {author} {\bibfnamefont {X.}~\bibnamefont {Pan}}, \bibinfo {author}
  {\bibfnamefont {X.}~\bibnamefont {Shen}}, \bibinfo {author} {\bibfnamefont
  {X.}~\bibnamefont {Guan}}, \bibinfo {author} {\bibfnamefont {D.}~\bibnamefont
  {Dai}}, \bibinfo {author} {\bibfnamefont {K.}~\bibnamefont {Zhong}},\ and\
  \bibinfo {author} {\bibfnamefont {L.}~\bibnamefont {Liu}},\ }\href
  {https://doi.org/10.1038/s41377-024-01506-1} {\bibfield  {journal} {\bibinfo
  {journal} {Light: Science \& Applications}\ }\textbf {\bibinfo {volume}
  {13}},\ \bibinfo {pages} {147} (\bibinfo {year} {2024})}\BibitemShut
  {NoStop}%
\bibitem [{\citenamefont {Chen}\ \emph {et~al.}(2022)\citenamefont {Chen},
  \citenamefont {Chen}, \citenamefont {Ruan}, \citenamefont {Fan},
  \citenamefont {Zhang}, \citenamefont {Gan}, \citenamefont {Liu},
  \citenamefont {Dai}, \citenamefont {Guo},\ and\ \citenamefont
  {Liu}}]{chen_cwdm_2022}%
  \BibitemOpen
  \bibfield  {author} {\bibinfo {author} {\bibfnamefont {K.}~\bibnamefont
  {Chen}}, \bibinfo {author} {\bibfnamefont {G.}~\bibnamefont {Chen}}, \bibinfo
  {author} {\bibfnamefont {Z.}~\bibnamefont {Ruan}}, \bibinfo {author}
  {\bibfnamefont {X.}~\bibnamefont {Fan}}, \bibinfo {author} {\bibfnamefont
  {J.}~\bibnamefont {Zhang}}, \bibinfo {author} {\bibfnamefont
  {R.}~\bibnamefont {Gan}}, \bibinfo {author} {\bibfnamefont {J.}~\bibnamefont
  {Liu}}, \bibinfo {author} {\bibfnamefont {D.}~\bibnamefont {Dai}}, \bibinfo
  {author} {\bibfnamefont {C.}~\bibnamefont {Guo}},\ and\ \bibinfo {author}
  {\bibfnamefont {L.}~\bibnamefont {Liu}},\ }\href
  {https://doi.org/10.1088/1674-4926/43/11/112301} {\bibfield  {journal}
  {\bibinfo  {journal} {Journal of Semiconductors}\ }\textbf {\bibinfo {volume}
  {43}},\ \bibinfo {pages} {112301} (\bibinfo {year} {2022})}\BibitemShut
  {NoStop}%
\bibitem [{\citenamefont {He}\ \emph {et~al.}(2025)\citenamefont {He},
  \citenamefont {Liu}, \citenamefont {Zhang}, \citenamefont {Liu},
  \citenamefont {Bao}, \citenamefont {Li}, \citenamefont {Zhao}, \citenamefont
  {Pan}, \citenamefont {Huang}, \citenamefont {Yu}, \citenamefont {Liu},
  \citenamefont {Shi},\ and\ \citenamefont {Dai}}]{he_athermal_pon_2025}%
  \BibitemOpen
  \bibfield  {author} {\bibinfo {author} {\bibfnamefont {J.}~\bibnamefont
  {He}}, \bibinfo {author} {\bibfnamefont {D.}~\bibnamefont {Liu}}, \bibinfo
  {author} {\bibfnamefont {M.}~\bibnamefont {Zhang}}, \bibinfo {author}
  {\bibfnamefont {H.}~\bibnamefont {Liu}}, \bibinfo {author} {\bibfnamefont
  {Y.}~\bibnamefont {Bao}}, \bibinfo {author} {\bibfnamefont {C.}~\bibnamefont
  {Li}}, \bibinfo {author} {\bibfnamefont {W.}~\bibnamefont {Zhao}}, \bibinfo
  {author} {\bibfnamefont {B.}~\bibnamefont {Pan}}, \bibinfo {author}
  {\bibfnamefont {Y.}~\bibnamefont {Huang}}, \bibinfo {author} {\bibfnamefont
  {Z.}~\bibnamefont {Yu}}, \bibinfo {author} {\bibfnamefont {L.}~\bibnamefont
  {Liu}}, \bibinfo {author} {\bibfnamefont {Y.}~\bibnamefont {Shi}},\ and\
  \bibinfo {author} {\bibfnamefont {D.}~\bibnamefont {Dai}},\ }\href
  {https://doi.org/10.1002/lpor.202401330} {\bibfield  {journal} {\bibinfo
  {journal} {Laser \& Photonics Reviews}\ }\textbf {\bibinfo {volume} {19}},\
  \bibinfo {pages} {2401330} (\bibinfo {year} {2025})}\BibitemShut {NoStop}%
\bibitem [{\citenamefont {Zhao}\ \emph {et~al.}(2025)\citenamefont {Zhao},
  \citenamefont {Zhu}, \citenamefont {Yi}, \citenamefont {Liu}, \citenamefont
  {Cao}, \citenamefont {Wang}, \citenamefont {Yan}, \citenamefont {Huang},
  \citenamefont {Yu},\ and\ \citenamefont
  {Dai}}]{zhao_modedivisionmultiplexing_2025}%
  \BibitemOpen
  \bibfield  {author} {\bibinfo {author} {\bibfnamefont {W.}~\bibnamefont
  {Zhao}}, \bibinfo {author} {\bibfnamefont {M.}~\bibnamefont {Zhu}}, \bibinfo
  {author} {\bibfnamefont {X.}~\bibnamefont {Yi}}, \bibinfo {author}
  {\bibfnamefont {H.}~\bibnamefont {Liu}}, \bibinfo {author} {\bibfnamefont
  {H.}~\bibnamefont {Cao}}, \bibinfo {author} {\bibfnamefont {S.}~\bibnamefont
  {Wang}}, \bibinfo {author} {\bibfnamefont {H.}~\bibnamefont {Yan}}, \bibinfo
  {author} {\bibfnamefont {F.}~\bibnamefont {Huang}}, \bibinfo {author}
  {\bibfnamefont {Z.}~\bibnamefont {Yu}},\ and\ \bibinfo {author}
  {\bibfnamefont {D.}~\bibnamefont {Dai}},\ }\href
  {https://doi.org/10.1002/lpor.202400861} {\bibfield  {journal} {\bibinfo
  {journal} {Laser \& Photonics Reviews}\ }\textbf {\bibinfo {volume} {19}},\
  \bibinfo {pages} {2400861} (\bibinfo {year} {2025})}\BibitemShut {NoStop}%
\bibitem [{\citenamefont {Hulyal}\ \emph {et~al.}(2025)\citenamefont {Hulyal},
  \citenamefont {Hu}, \citenamefont {Wang}, \citenamefont {Cai}, \citenamefont
  {Lihachev},\ and\ \citenamefont {Kippenberg}}]{hulyal_arrayed_2025}%
  \BibitemOpen
  \bibfield  {author} {\bibinfo {author} {\bibfnamefont {S.~U.}\ \bibnamefont
  {Hulyal}}, \bibinfo {author} {\bibfnamefont {J.}~\bibnamefont {Hu}}, \bibinfo
  {author} {\bibfnamefont {C.}~\bibnamefont {Wang}}, \bibinfo {author}
  {\bibfnamefont {J.}~\bibnamefont {Cai}}, \bibinfo {author} {\bibfnamefont
  {G.}~\bibnamefont {Lihachev}},\ and\ \bibinfo {author} {\bibfnamefont
  {T.~J.}\ \bibnamefont {Kippenberg}},\ }\href
  {https://doi.org/10.1364/OPTICA.565570} {\bibfield  {journal} {\bibinfo
  {journal} {Optica}\ }\textbf {\bibinfo {volume} {12}},\ \bibinfo {pages}
  {978} (\bibinfo {year} {2025})}\BibitemShut {NoStop}%
\bibitem [{\citenamefont {Wang}\ \emph
  {et~al.}(2024{\natexlab{a}})\citenamefont {Wang}, \citenamefont {Fang},
  \citenamefont {Zhang}, \citenamefont {Kotz}, \citenamefont {Lihachev},
  \citenamefont {Churaev}, \citenamefont {Li}, \citenamefont {Schwarzenberger},
  \citenamefont {Ou}, \citenamefont {Koos},\ and\ \citenamefont
  {Kippenberg}}]{wang_ultrabroadband_2024}%
  \BibitemOpen
  \bibfield  {author} {\bibinfo {author} {\bibfnamefont {C.}~\bibnamefont
  {Wang}}, \bibinfo {author} {\bibfnamefont {D.}~\bibnamefont {Fang}}, \bibinfo
  {author} {\bibfnamefont {J.}~\bibnamefont {Zhang}}, \bibinfo {author}
  {\bibfnamefont {A.}~\bibnamefont {Kotz}}, \bibinfo {author} {\bibfnamefont
  {G.}~\bibnamefont {Lihachev}}, \bibinfo {author} {\bibfnamefont
  {M.}~\bibnamefont {Churaev}}, \bibinfo {author} {\bibfnamefont
  {Z.}~\bibnamefont {Li}}, \bibinfo {author} {\bibfnamefont {A.}~\bibnamefont
  {Schwarzenberger}}, \bibinfo {author} {\bibfnamefont {X.}~\bibnamefont {Ou}},
  \bibinfo {author} {\bibfnamefont {C.}~\bibnamefont {Koos}},\ and\ \bibinfo
  {author} {\bibfnamefont {T.~J.}\ \bibnamefont {Kippenberg}},\ }\href
  {https://doi.org/10.1364/OPTICA.537730} {\bibfield  {journal} {\bibinfo
  {journal} {Optica}\ }\textbf {\bibinfo {volume} {11}},\ \bibinfo {pages}
  {1614} (\bibinfo {year} {2024}{\natexlab{a}})}\BibitemShut {NoStop}%
\bibitem [{\citenamefont {Lin}\ \emph {et~al.}(2026)\citenamefont {Lin},
  \citenamefont {Li}, \citenamefont {Kotz}, \citenamefont {Larocque},
  \citenamefont {Kuznetsov}, \citenamefont {Sun}, \citenamefont {Zhang},
  \citenamefont {Zheng}, \citenamefont {Riemensberger}, \citenamefont {Koos},\
  and\ \citenamefont {Kippenberg}}]{lin_copper_2026}%
  \BibitemOpen
  \bibfield  {author} {\bibinfo {author} {\bibfnamefont {M.}~\bibnamefont
  {Lin}}, \bibinfo {author} {\bibfnamefont {Z.}~\bibnamefont {Li}}, \bibinfo
  {author} {\bibfnamefont {A.}~\bibnamefont {Kotz}}, \bibinfo {author}
  {\bibfnamefont {H.}~\bibnamefont {Larocque}}, \bibinfo {author}
  {\bibfnamefont {N.}~\bibnamefont {Kuznetsov}}, \bibinfo {author}
  {\bibfnamefont {J.}~\bibnamefont {Sun}}, \bibinfo {author} {\bibfnamefont
  {Y.}~\bibnamefont {Zhang}}, \bibinfo {author} {\bibfnamefont
  {S.}~\bibnamefont {Zheng}}, \bibinfo {author} {\bibfnamefont
  {J.}~\bibnamefont {Riemensberger}}, \bibinfo {author} {\bibfnamefont
  {C.}~\bibnamefont {Koos}},\ and\ \bibinfo {author} {\bibfnamefont {T.~J.}\
  \bibnamefont {Kippenberg}},\ }\href
  {https://doi.org/10.1038/s41467-026-69588-6} {\bibfield  {journal} {\bibinfo
  {journal} {Nature Communications}\ }\textbf {\bibinfo {volume} {17}},\
  \bibinfo {pages} {3211} (\bibinfo {year} {2026})}\BibitemShut {NoStop}%
\bibitem [{\citenamefont {Smit}\ and\ \citenamefont
  {Van~Dam}(1996)}]{smit_phasar-based_1996}%
  \BibitemOpen
  \bibfield  {author} {\bibinfo {author} {\bibfnamefont {M.~K.}\ \bibnamefont
  {Smit}}\ and\ \bibinfo {author} {\bibfnamefont {C.}~\bibnamefont {Van~Dam}},\
  }\href {https://doi.org/10.1109/2944.577370} {\bibfield  {journal} {\bibinfo
  {journal} {IEEE Journal of Selected Topics in Quantum Electronics}\ }\textbf
  {\bibinfo {volume} {2}},\ \bibinfo {pages} {236} (\bibinfo {year}
  {1996})}\BibitemShut {NoStop}%
\bibitem [{\citenamefont {Hulyal}\ \emph {et~al.}(2026)\citenamefont {Hulyal},
  \citenamefont {Fang}, \citenamefont {Cai}, \citenamefont {Kotz},
  \citenamefont {Ou}, \citenamefont {Koos},\ and\ \citenamefont
  {Kippenberg}}]{hulyal_16_2026}%
  \BibitemOpen
  \bibfield  {author} {\bibinfo {author} {\bibfnamefont {S.~U.}\ \bibnamefont
  {Hulyal}}, \bibinfo {author} {\bibfnamefont {D.}~\bibnamefont {Fang}},
  \bibinfo {author} {\bibfnamefont {J.}~\bibnamefont {Cai}}, \bibinfo {author}
  {\bibfnamefont {A.}~\bibnamefont {Kotz}}, \bibinfo {author} {\bibfnamefont
  {X.}~\bibnamefont {Ou}}, \bibinfo {author} {\bibfnamefont {C.}~\bibnamefont
  {Koos}},\ and\ \bibinfo {author} {\bibfnamefont {T.~J.}\ \bibnamefont
  {Kippenberg}},\ }in\ \href {https://doi.org/10.1364/OFC.2026.W1A.7} {\emph
  {\bibinfo {booktitle} {Optical Fiber Communication Conference (OFC) 2026}}}\
  (\bibinfo {year} {2026})\ p.\ \bibinfo {pages} {W1A.7}\BibitemShut {NoStop}%
\bibitem [{\citenamefont {Wang}\ \emph
  {et~al.}(2024{\natexlab{b}})\citenamefont {Wang}, \citenamefont {Li},
  \citenamefont {Riemensberger}, \citenamefont {Lihachev}, \citenamefont
  {Churaev}, \citenamefont {Kao}, \citenamefont {Ji}, \citenamefont {Zhang},
  \citenamefont {Blesin}, \citenamefont {Davydova}, \citenamefont {Chen},
  \citenamefont {Huang}, \citenamefont {Wang}, \citenamefont {Ou},\ and\
  \citenamefont {Kippenberg}}]{wang_lithium_2024}%
  \BibitemOpen
  \bibfield  {author} {\bibinfo {author} {\bibfnamefont {C.}~\bibnamefont
  {Wang}}, \bibinfo {author} {\bibfnamefont {Z.}~\bibnamefont {Li}}, \bibinfo
  {author} {\bibfnamefont {J.}~\bibnamefont {Riemensberger}}, \bibinfo {author}
  {\bibfnamefont {G.}~\bibnamefont {Lihachev}}, \bibinfo {author}
  {\bibfnamefont {M.}~\bibnamefont {Churaev}}, \bibinfo {author} {\bibfnamefont
  {W.}~\bibnamefont {Kao}}, \bibinfo {author} {\bibfnamefont {X.}~\bibnamefont
  {Ji}}, \bibinfo {author} {\bibfnamefont {J.}~\bibnamefont {Zhang}}, \bibinfo
  {author} {\bibfnamefont {T.}~\bibnamefont {Blesin}}, \bibinfo {author}
  {\bibfnamefont {A.}~\bibnamefont {Davydova}}, \bibinfo {author}
  {\bibfnamefont {Y.}~\bibnamefont {Chen}}, \bibinfo {author} {\bibfnamefont
  {K.}~\bibnamefont {Huang}}, \bibinfo {author} {\bibfnamefont
  {X.}~\bibnamefont {Wang}}, \bibinfo {author} {\bibfnamefont {X.}~\bibnamefont
  {Ou}},\ and\ \bibinfo {author} {\bibfnamefont {T.~J.}\ \bibnamefont
  {Kippenberg}},\ }\href {https://doi.org/10.1038/s41586-024-07369-1}
  {\bibfield  {journal} {\bibinfo  {journal} {Nature}\ }\textbf {\bibinfo
  {volume} {629}},\ \bibinfo {pages} {784} (\bibinfo {year}
  {2024}{\natexlab{b}})}\BibitemShut {NoStop}%
\end{thebibliography}%


\begin{thebibliography}{5}%
\makeatletter
\providecommand \@ifxundefined [1]{%
 \@ifx{#1\undefined}
}%
\providecommand \@ifnum [1]{%
 \ifnum #1\expandafter \@firstoftwo
 \else \expandafter \@secondoftwo
 \fi
}%
\providecommand \@ifx [1]{%
 \ifx #1\expandafter \@firstoftwo
 \else \expandafter \@secondoftwo
 \fi
}%
\providecommand \natexlab [1]{#1}%
\providecommand \enquote  [1]{``#1''}%
\providecommand \bibnamefont  [1]{#1}%
\providecommand \bibfnamefont [1]{#1}%
\providecommand \citenamefont [1]{#1}%
\providecommand \href@noop [0]{\@secondoftwo}%
\providecommand \href [0]{\begingroup \@sanitize@url \@href}%
\providecommand \@href[1]{\@@startlink{#1}\@@href}%
\providecommand \@@href[1]{\endgroup#1\@@endlink}%
\providecommand \@sanitize@url [0]{\catcode `\\12\catcode `\$12\catcode
  `\&12\catcode `\#12\catcode `\^12\catcode `\_12\catcode `\%12\relax}%
\providecommand \@@startlink[1]{}%
\providecommand \@@endlink[0]{}%
\providecommand \url  [0]{\begingroup\@sanitize@url \@url }%
\providecommand \@url [1]{\endgroup\@href {#1}{\urlprefix }}%
\providecommand \urlprefix  [0]{URL }%
\providecommand \Eprint [0]{\href }%
\providecommand \doibase [0]{https://doi.org/}%
\providecommand \selectlanguage [0]{\@gobble}%
\providecommand \bibinfo  [0]{\@secondoftwo}%
\providecommand \bibfield  [0]{\@secondoftwo}%
\providecommand \translation [1]{[#1]}%
\providecommand \BibitemOpen [0]{}%
\providecommand \bibitemStop [0]{}%
\providecommand \bibitemNoStop [0]{.\EOS\space}%
\providecommand \EOS [0]{\spacefactor3000\relax}%
\providecommand \BibitemShut  [1]{\csname bibitem#1\endcsname}%
\let\auto@bib@innerbib\@empty
\bibitem [{\citenamefont {Smit}\ and\ \citenamefont
  {Van~Dam}(1996)}]{smit_phasar-based_1996}%
  \BibitemOpen
  \bibfield  {author} {\bibinfo {author} {\bibfnamefont {M.~K.}\ \bibnamefont
  {Smit}}\ and\ \bibinfo {author} {\bibfnamefont {C.}~\bibnamefont {Van~Dam}},\
  }\href {https://doi.org/10.1109/2944.577370} {\bibfield  {journal} {\bibinfo
  {journal} {IEEE Journal of Selected Topics in Quantum Electronics}\ }\textbf
  {\bibinfo {volume} {2}},\ \bibinfo {pages} {236} (\bibinfo {year}
  {1996})}\BibitemShut {NoStop}%
\bibitem [{\citenamefont {Hulyal}\ \emph {et~al.}(2025)\citenamefont {Hulyal},
  \citenamefont {Hu}, \citenamefont {Wang}, \citenamefont {Cai}, \citenamefont
  {Lihachev},\ and\ \citenamefont {Kippenberg}}]{hulyal_arrayed_2025}%
  \BibitemOpen
  \bibfield  {author} {\bibinfo {author} {\bibfnamefont {S.~U.}\ \bibnamefont
  {Hulyal}}, \bibinfo {author} {\bibfnamefont {J.}~\bibnamefont {Hu}}, \bibinfo
  {author} {\bibfnamefont {C.}~\bibnamefont {Wang}}, \bibinfo {author}
  {\bibfnamefont {J.}~\bibnamefont {Cai}}, \bibinfo {author} {\bibfnamefont
  {G.}~\bibnamefont {Lihachev}},\ and\ \bibinfo {author} {\bibfnamefont
  {T.~J.}\ \bibnamefont {Kippenberg}},\ }\href
  {https://doi.org/10.1364/OPTICA.565570} {\bibfield  {journal} {\bibinfo
  {journal} {Optica}\ }\textbf {\bibinfo {volume} {12}},\ \bibinfo {pages}
  {978} (\bibinfo {year} {2025})}\BibitemShut {NoStop}%
\bibitem [{\citenamefont {Hu}\ \emph {et~al.}(2022)\citenamefont {Hu},
  \citenamefont {Borkowski}, \citenamefont {Lefevre}, \citenamefont {Cho},
  \citenamefont {Buchali}, \citenamefont {Bonk}, \citenamefont {Schuh},
  \citenamefont {De~Leo}, \citenamefont {Habegger}, \citenamefont {Destraz},
  \citenamefont {Del~Medico}, \citenamefont {Duran}, \citenamefont {Tedaldi},
  \citenamefont {Funck}, \citenamefont {Fedoryshyn}, \citenamefont {Leuthold},
  \citenamefont {Heni}, \citenamefont {Baeuerle},\ and\ \citenamefont
  {Hoessbacher}}]{hu_ultrahigh-net-bitrate_2022}%
  \BibitemOpen
  \bibfield  {author} {\bibinfo {author} {\bibfnamefont {Q.}~\bibnamefont
  {Hu}}, \bibinfo {author} {\bibfnamefont {R.}~\bibnamefont {Borkowski}},
  \bibinfo {author} {\bibfnamefont {Y.}~\bibnamefont {Lefevre}}, \bibinfo
  {author} {\bibfnamefont {J.}~\bibnamefont {Cho}}, \bibinfo {author}
  {\bibfnamefont {F.}~\bibnamefont {Buchali}}, \bibinfo {author} {\bibfnamefont
  {R.}~\bibnamefont {Bonk}}, \bibinfo {author} {\bibfnamefont {K.}~\bibnamefont
  {Schuh}}, \bibinfo {author} {\bibfnamefont {E.}~\bibnamefont {De~Leo}},
  \bibinfo {author} {\bibfnamefont {P.}~\bibnamefont {Habegger}}, \bibinfo
  {author} {\bibfnamefont {M.}~\bibnamefont {Destraz}}, \bibinfo {author}
  {\bibfnamefont {N.}~\bibnamefont {Del~Medico}}, \bibinfo {author}
  {\bibfnamefont {H.}~\bibnamefont {Duran}}, \bibinfo {author} {\bibfnamefont
  {V.}~\bibnamefont {Tedaldi}}, \bibinfo {author} {\bibfnamefont
  {C.}~\bibnamefont {Funck}}, \bibinfo {author} {\bibfnamefont
  {Y.}~\bibnamefont {Fedoryshyn}}, \bibinfo {author} {\bibfnamefont
  {J.}~\bibnamefont {Leuthold}}, \bibinfo {author} {\bibfnamefont
  {W.}~\bibnamefont {Heni}}, \bibinfo {author} {\bibfnamefont {B.}~\bibnamefont
  {Baeuerle}},\ and\ \bibinfo {author} {\bibfnamefont {C.}~\bibnamefont
  {Hoessbacher}},\ }\href {https://doi.org/10.1109/JLT.2022.3172246} {\bibfield
   {journal} {\bibinfo  {journal} {Journal of Lightwave Technology}\ }\textbf
  {\bibinfo {volume} {40}},\ \bibinfo {pages} {3338} (\bibinfo {year}
  {2022})}\BibitemShut {NoStop}%
\bibitem [{\citenamefont {Ji}\ \emph {et~al.}(2025)\citenamefont {Ji},
  \citenamefont {Li}, \citenamefont {Qiu}, \citenamefont {Wang}, \citenamefont
  {Divall}, \citenamefont {Gelash}, \citenamefont {Lihachev},\ and\
  \citenamefont {Kippenberg}}]{ji_deterministic_2025}%
  \BibitemOpen
  \bibfield  {author} {\bibinfo {author} {\bibfnamefont {X.}~\bibnamefont
  {Ji}}, \bibinfo {author} {\bibfnamefont {X.}~\bibnamefont {Li}}, \bibinfo
  {author} {\bibfnamefont {Z.}~\bibnamefont {Qiu}}, \bibinfo {author}
  {\bibfnamefont {R.~N.}\ \bibnamefont {Wang}}, \bibinfo {author}
  {\bibfnamefont {M.}~\bibnamefont {Divall}}, \bibinfo {author} {\bibfnamefont
  {A.}~\bibnamefont {Gelash}}, \bibinfo {author} {\bibfnamefont
  {G.}~\bibnamefont {Lihachev}},\ and\ \bibinfo {author} {\bibfnamefont
  {T.~J.}\ \bibnamefont {Kippenberg}},\ }\href
  {https://doi.org/10.1038/s41586-025-09598-4} {\bibfield  {journal} {\bibinfo
  {journal} {Nature}\ }\textbf {\bibinfo {volume} {646}},\ \bibinfo {pages}
  {843} (\bibinfo {year} {2025})}\BibitemShut {NoStop}%
\bibitem [{\citenamefont {Herr}\ \emph {et~al.}(2014)\citenamefont {Herr},
  \citenamefont {Brasch}, \citenamefont {Jost}, \citenamefont {Wang},
  \citenamefont {Kondratiev}, \citenamefont {Gorodetsky},\ and\ \citenamefont
  {Kippenberg}}]{herr_temporal_2014}%
  \BibitemOpen
  \bibfield  {author} {\bibinfo {author} {\bibfnamefont {T.}~\bibnamefont
  {Herr}}, \bibinfo {author} {\bibfnamefont {V.}~\bibnamefont {Brasch}},
  \bibinfo {author} {\bibfnamefont {J.~D.}\ \bibnamefont {Jost}}, \bibinfo
  {author} {\bibfnamefont {C.~Y.}\ \bibnamefont {Wang}}, \bibinfo {author}
  {\bibfnamefont {N.~M.}\ \bibnamefont {Kondratiev}}, \bibinfo {author}
  {\bibfnamefont {M.~L.}\ \bibnamefont {Gorodetsky}},\ and\ \bibinfo {author}
  {\bibfnamefont {T.~J.}\ \bibnamefont {Kippenberg}},\ }\href
  {https://doi.org/10.1038/nphoton.2013.343} {\bibfield  {journal} {\bibinfo
  {journal} {Nature Photonics}\ }\textbf {\bibinfo {volume} {8}},\ \bibinfo
  {pages} {145} (\bibinfo {year} {2014})}\BibitemShut {NoStop}%
\end{thebibliography}%
	\bibliographystyle{apsrev4-2}

\end{document}


\title{Supplementary Information for: Terabit-scale wavelength-multiplexed transmitter enabled by thin-film lithium tantalate}

\author{Shivaprasad U. Hulyal$^{1,2,*}$, Dengyang Fang$^{3,*}$, Jiachen Cai$^{1,2,*}$, Xinru Ji$^{1,2}$, Alexander Kotz$^{3}$, Jiale Sun$^{1,2}$, Christian Koos$^{3,\dagger}$, and Tobias J. Kippenberg$^{1,2,\ddagger}$}

\affiliation{$^1$Institute of Physics, Swiss Federal Institute of Technology Lausanne (EPFL), CH-1015 Lausanne, Switzerland\\
$^2$Institute of Electrical and Micro Engineering, Swiss Federal Institute of Technology Lausanne (EPFL), CH-1015 Lausanne, Switzerland\\
$^3$Institute of Photonics and Quantum Electronics (IPQ), Karlsruhe Institute of Technology (KIT), 76131 Karlsruhe, Germany\\
$^*$Shivaprasad U. Hulyal, Dengyang Fang and Jiachen Cai contributed equally to this work.\\
$^\dagger$christian.koos@kit.edu\quad $^\ddagger$tobias.kippenberg@epfl.ch
}

\setcounter{equation}{0}
\setcounter{figure}{0}
\setcounter{table}{0}
\setcounter{subsection}{0}
\setcounter{section}{0}
\setcounter{secnumdepth}{3}

\maketitle
{\hypersetup{linkcolor=blue}\tableofcontents}
\newpage

\section{Arrayed waveguide grating building block}
The arrayed waveguide gratings (AWGs) use a Rowland star-coupler configuration~\cite{smit_phasar-based_1996}, with diffraction order $m=160$, $N=13$ arrayed waveguides, path-length increment $\Delta L\approx129.0\,\mu$m and grating-circle radius $R_a=75\,\mu$m. The layout was developed using the semi-analytic diffraction model and validated by the finite-difference time-domain simulations reported in Ref.~\cite{hulyal_arrayed_2025}. The integrated transmitter uses this layout and was fabricated in a separate run on wafer D210. The fabrication process is described in the Methods section of the main text.

\section{Integrated active transmitter and cascaded passband}
\label{sec:device}
For an unmodulated carrier, the power transmission through channel $j$ is proportional to the product of the demultiplexer and multiplexer power responses. We use the standalone-AWG measurement in main text Fig.~2j as a reference for each AWG response. Omitting the bias-dependent transmission of the Mach--Zehnder modulator (MZM) and other approximately frequency-independent losses gives
\begin{equation}
T_j(\nu) \;=\; T^{\mathrm{demux}}_j(\nu)\,\cdot\,T^{\mathrm{mux}}_j(\nu).
\label{eq:cascade}
\end{equation}
Both $T_j^{\mathrm{demux}}$ and $T_j^{\mathrm{mux}}$ are power-transmission functions. For equal Gaussian power passbands with individual full width at half maximum $B_1$, the normalized product has width
\begin{equation}
B_2 \;=\; \frac{B_1}{\sqrt{2}} \;\approx\; 0.707\,B_1,
\label{eq:gauss}
\end{equation}
which gives $B_2=59.7$\,GHz for $B_1=84.4$\,GHz. If the two Gaussian centres differ by $\delta\nu$, their product can be written as
\begin{equation}
T_j(\nu)=A_j\exp\!\left[-2\ln2\left(\frac{\delta\nu}{B_1}\right)^2\right]
\exp\!\left[-8\ln2\left(\frac{\nu-\bar\nu_j}{B_1}\right)^2\right],
\label{eq:gauss_offset}
\end{equation}
where $\bar\nu_j$ is the mean of the two centre frequencies and $A_j$ is the product of their peak transmissions. Within this model, the offset reduces the peak transmission but leaves the normalized product width unchanged. Non-Gaussian responses or unequal individual widths can modify the measured product shape.

\begin{figure}[!htbp]
\centering
\includegraphics[width=\suppfigurewidth]{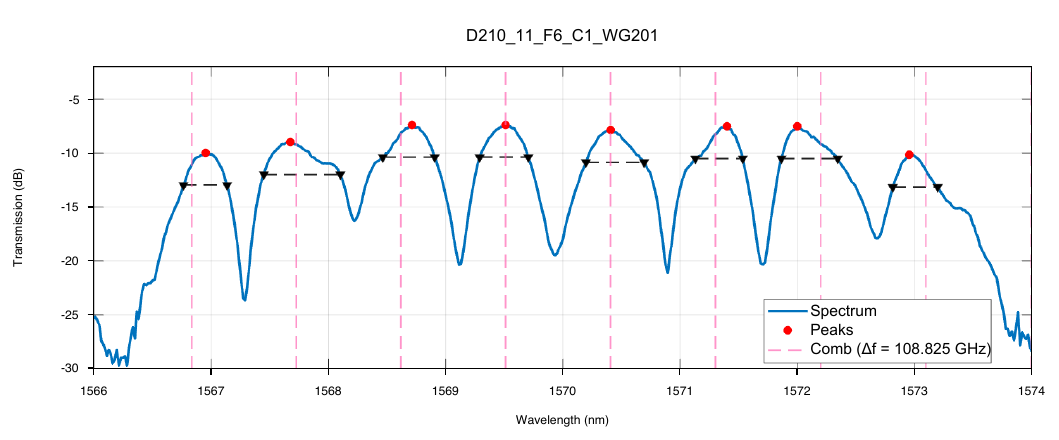}
\caption{\textbf{Swept-carrier transmission of the integrated transmitter.} Device D210\_11\_F6\_C1\_WG201 was measured over $1566$--$1574$\,nm using comb-assisted frequency calibration. Red markers identify the eight peaks, and horizontal bars show their $3$\,dB widths, averaging $55.3$\,GHz. Dashed lines mark the $108.825$\,GHz reference grid of the sweep, distinct from the $106.2$\,GHz mean spacing of the standalone AWGs.}
\label{fig:passband}
\end{figure}

The swept-carrier measurement in Fig.~\ref{fig:passband} covers a frequency interval distinct from the carrier grid used in the transmission experiments of Sec.~\ref{sec:tx}. The sweep records the combined AWG response at a fixed MZM bias, and the measured peak transmissions span a $5.96$\,dB range. If the MZM transmission is approximately constant across each passband, it changes the peak level without changing the normalized width.

During data transmission, the MZM generates modulation sidebands between the two AWGs, so these sidebands pass through only the output multiplexer. At carrier frequency $\nu_j$, the AWG contribution to the modulated field response at frequency offset $f$ is proportional to $H_j^{\mathrm{demux}}(\nu_j)H_j^{\mathrm{mux}}(\nu_j+f)$, where $H$ denotes a field transfer function. The distinction between the passive cascade and filtering of modulation sidebands is discussed in the Passive and electro-optic characterization section of the main text.

\section{Modulator electrode design}
\label{sec:mzm}
The modulators use $7.6$\,mm-long gold travelling-wave electrodes in a ground--signal--ground configuration, with a $2.25\,\mu$m electrode-to-waveguide gap. The design targets microwave--optical velocity matching and impedance matching. The waveguide cross-section and DC transfer curve are shown in main text Fig.~2c and i, respectively. The transmission measurement procedure is described in Sec.~\ref{sec:tx}.

\section{High-speed transmission experiment}
\label{sec:tx}
\subsection{Experimental setup}
Two continuous-wave carriers from external-cavity lasers (ECLs) are tuned to the addressed WDM channels, combined in a polarization-maintaining coupler and launched into the \LiTaO{} transmitter with aligned polarization (main text Fig.~3a). Two digital-to-analogue converters (DACs, Keysight M8199B, $256$\,GSa\,s$^{-1}$) generate the electrical drive signals, which are amplified by RF amplifiers (Anritsu AH15199B) and applied to adjacent MZMs through a $110$\,GHz GSGGSG probe. A second probe, connected to two bias-Ts and $50\,\Omega$ terminations, sets the quadrature biases and terminates the travelling-wave electrodes. Digital root-raised-cosine filters shape the transmitted pulses, with roll-off factors of $0.01$ and $0.05$ for PAM4 and PAM8, respectively. The adjacent channels carry independent data simultaneously.

An erbium-doped fibre amplifier (EDFA) with a noise figure of approximately $5$\,dB amplifies the multiplexed output. A $5\%$ tap feeds an optical spectrum analyser (OSA, Finisar WaveAnalyzer), while the remaining $95\%$ goes to the receiver. A programmable optical demultiplexer (Finisar WaveShaper) selects each channel and suppresses out-of-band amplified spontaneous-emission (ASE) noise. A $90$\,GHz single-ended photodiode (Finisar BPDV4120R) detects each channel sequentially, and a real-time oscilloscope (Keysight UXR 1004A) acquires the waveform for offline processing.

\subsection{Digital signal processing and rate definitions}
The receiver digital signal processing (DSP) comprises resampling to two samples per symbol, timing recovery, linear Sato equalization, decision-directed least-mean-square (DD-LMS) equalization and symbol demapping. The recovered symbols are used to evaluate the bit-error ratio (BER) before forward-error correction (FEC), channel signal-to-noise ratio (CSNR) and generalized mutual information (GMI). For symbol rate $B_{\mathrm{sym}}$, the achievable information rate (AIR) is
\begin{equation}
\mathrm{AIR}=\mathrm{GMI}\,B_{\mathrm{sym}},
\label{eq:air}
\end{equation}
where GMI is expressed in bit per symbol. The gross line rate is $B_{\mathrm{sym}}\log_2 M$ for an $M$-level PAM alphabet. GMI-based rate assessment follows the framework of Ref.~\cite{hu_ultrahigh-net-bitrate_2022}.

The net data rates (NDRs) in Tables~\ref{tab:pam4}, \ref{tab:pam8} and \ref{tab:comb} are estimates from the measurement analysis. The pre-FEC BER reference criteria used in that analysis are $4\times10^{-2}$ for the laser-fed WDM measurements and $2\times10^{-2}$ for the carrier-power sweep, associated with $20\%$ and $15\%$ soft-decision FEC overhead, respectively.

\subsection{Laser-fed transmission results}
\begin{figure}[!htbp]
\centering
\includegraphics[width=\suppfigurewidth]{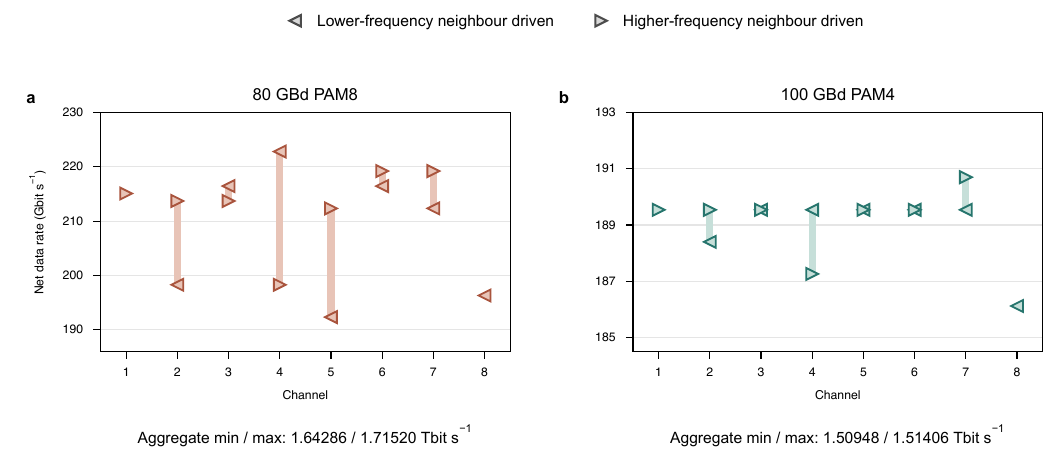}
\caption{\textbf{Estimated net data rates for adjacent-pair measurements.} \textbf{(a)} $80$\,GBd PAM8. \textbf{(b)} $100$\,GBd PAM4. Left- and right-pointing markers identify the driven lower- and higher-frequency neighbour, respectively. Connectors join the two records of each interior channel. Each panel shows the minimum- and maximum-record aggregates from Tables~\ref{tab:pam8} and \ref{tab:pam4}. The vertical scales differ.}
\label{fig:pairwise}
\end{figure}

The eight laser-fed channels span $192.517$--$193.259$\,THz on a nominal $106$\,GHz grid. Measurements were acquired for seven adjacent-pair configurations: $1/2$, $2/3$, $3/4$, $4/5$, $5/6$, $6/7$ and $7/8$. Each interior channel therefore has two measurement records, one with its lower-frequency neighbour driven and one with its higher-frequency neighbour driven. Tables~\ref{tab:pam4} and \ref{tab:pam8} list all 14 records for each format. The variation in estimated NDR between pair configurations is shown in Fig.~\ref{fig:pairwise}a for PAM8 and Fig.~\ref{fig:pairwise}b for PAM4.

\begin{figure}[!htbp]
\centering
\includegraphics[width=\suppfigurewidth]{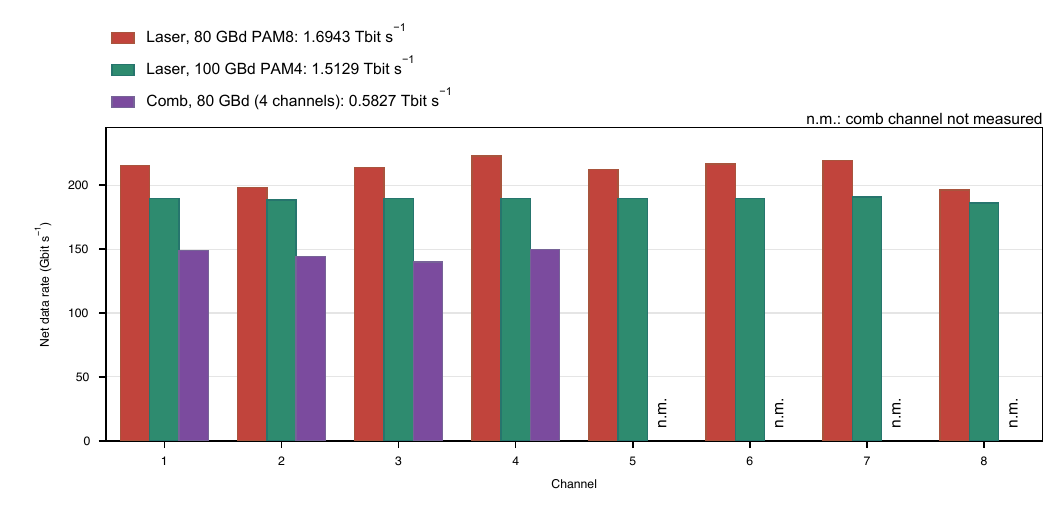}
\caption{\textbf{Per-channel estimated net data rates.} Laser-fed PAM8 and PAM4 rates use the disjoint-pair records ($1/2$, $3/4$, $5/6$ and $7/8$). The comb-driven series uses the four PAM4-equivalent records in Table~\ref{tab:comb}, with unmeasured channels marked. Legend totals sum the displayed records. Channel indices refer to each experiment's carrier grid.}
\label{fig:dashboard}
\end{figure}

The aggregate rows in the tables distinguish the disjoint-pair selection used in main text Fig.~3 ($1/2$, $3/4$, $5/6$ and $7/8$) from selections of the minimum or maximum NDR record for each channel. Figure~\ref{fig:dashboard} compares the estimated NDRs from the disjoint-pair measurements with those of the four comb-driven channels described in Sec.~\ref{sec:comb}.

\begin{table}[htbp]
\centering
\caption{\textbf{Laser-fed $100$\,GBd PAM4 transmission.} Each row records one channel while the indicated adjacent pair carries independent data simultaneously. AIR and estimated NDR are defined in the text.}
\label{tab:pam4}
\setlength{\tabcolsep}{5pt}
\begin{tabular}{cccccccc}
\toprule
\makecell{Pair\\(MZM)} & Channel & \makecell{Carrier\\(THz)} & BER & \makecell{CSNR\\(dB)} & \makecell{GMI\\(bit/sym)} & \makecell{AIR\\(Gbit/s)} & \makecell{NDR\\(Gbit/s)} \\
\midrule
1/2 & 1 & 192.517 & $7.3\times10^{-4}$ & 13.86 & 1.991 & 199.14 & 189.54 \\
1/2 & 2 & 192.623 & $1.1\times10^{-3}$ & 13.69 & 1.985 & 198.54 & 188.40 \\
2/3 & 2 & 192.623 & $1.9\times10^{-4}$ & 14.83 & 1.997 & 199.74 & 189.54 \\
2/3 & 3 & 192.729 & $3.1\times10^{-4}$ & 14.57 & 1.996 & 199.58 & 189.54 \\
3/4 & 3 & 192.729 & $9.9\times10^{-4}$ & 13.61 & 1.986 & 198.60 & 189.54 \\
3/4 & 4 & 192.835 & $1.1\times10^{-3}$ & 13.61 & 1.987 & 198.73 & 189.54 \\
4/5 & 4 & 192.835 & $3.0\times10^{-3}$ & 12.48 & 1.968 & 196.78 & 187.26 \\
4/5 & 5 & 192.941 & $1.0\times10^{-3}$ & 13.02 & 1.990 & 199.00 & 189.54 \\
5/6 & 5 & 192.941 & $4.3\times10^{-4}$ & 14.19 & 1.995 & 199.48 & 189.54 \\
5/6 & 6 & 193.047 & $1.1\times10^{-3}$ & 13.61 & 1.987 & 198.74 & 189.54 \\
6/7 & 6 & 193.047 & $2.1\times10^{-4}$ & 15.05 & 1.997 & 199.69 & 189.54 \\
6/7 & 7 & 193.153 & $2.3\times10^{-4}$ & 15.18 & 1.997 & 199.68 & 189.54 \\
7/8 & 7 & 193.153 & $1.3\times10^{-4}$ & 15.73 & 1.998 & 199.82 & 190.70 \\
7/8 & 8 & 193.259 & $4.5\times10^{-3}$ & 11.90 & 1.953 & 195.33 & 186.12 \\
\midrule
\multicolumn{7}{r}{Aggregate gross line rate\quad} & $1.600$\,Tbit/s \\
\multicolumn{7}{r}{Estimated NDR sum, minimum record per channel\quad} & $1.50948$\,Tbit/s \\
\multicolumn{7}{r}{Estimated NDR sum, disjoint-pair records\quad} & $1.51292$\,Tbit/s \\
\multicolumn{7}{r}{Estimated NDR sum, maximum record per channel\quad} & $1.51406$\,Tbit/s \\
\bottomrule
\end{tabular}
\end{table}

\begin{table}[htbp]
\centering
\caption{\textbf{Laser-fed $80$\,GBd PAM8 transmission.} Each row records one channel while the indicated adjacent pair carries independent data simultaneously. Disjoint-pair totals select configurations $1/2$, $3/4$, $5/6$ and $7/8$.}
\label{tab:pam8}
\setlength{\tabcolsep}{5pt}
\begin{tabular}{cccccccc}
\toprule
\makecell{Pair\\(MZM)} & Channel & \makecell{Carrier\\(THz)} & BER & \makecell{CSNR\\(dB)} & \makecell{GMI\\(bit/sym)} & \makecell{AIR\\(Gbit/s)} & \makecell{NDR\\(Gbit/s)} \\
\midrule
1/2 & 1 & 192.517 & $0.0111$ & 16.54 & 2.832 & 226.55 & 215.09 \\
1/2 & 2 & 192.623 & $0.0236$ & 15.07 & 2.648 & 211.81 & 198.29 \\
2/3 & 2 & 192.623 & $0.0114$ & 16.60 & 2.816 & 225.27 & 213.72 \\
2/3 & 3 & 192.729 & $0.0103$ & 16.70 & 2.844 & 227.48 & 216.46 \\
3/4 & 3 & 192.729 & $0.0120$ & 16.42 & 2.816 & 225.31 & 213.72 \\
3/4 & 4 & 192.835 & $0.0047$ & 18.02 & 2.908 & 232.62 & 222.82 \\
4/5 & 4 & 192.835 & $0.0255$ & 14.93 & 2.643 & 211.44 & 198.29 \\
4/5 & 5 & 192.941 & $0.0347$ & 14.10 & 2.562 & 204.97 & 192.34 \\
5/6 & 5 & 192.941 & $0.0145$ & 16.07 & 2.788 & 223.05 & 212.35 \\
5/6 & 6 & 193.047 & $0.0087$ & 17.10 & 2.844 & 227.48 & 216.46 \\
6/7 & 6 & 193.047 & $0.0072$ & 17.28 & 2.883 & 230.63 & 219.22 \\
6/7 & 7 & 193.153 & $0.0140$ & 16.17 & 2.785 & 222.83 & 212.35 \\
7/8 & 7 & 193.153 & $0.0071$ & 17.38 & 2.877 & 230.15 & 219.22 \\
7/8 & 8 & 193.259 & $0.0266$ & 14.78 & 2.615 & 209.21 & 196.32 \\
\midrule
\multicolumn{7}{r}{Aggregate gross line rate\quad} & $1.920$\,Tbit/s \\
\multicolumn{7}{r}{Estimated NDR sum, minimum record per channel\quad} & $1.64286$\,Tbit/s \\
\multicolumn{7}{r}{Estimated NDR sum, disjoint-pair records\quad} & $1.69427$\,Tbit/s \\
\multicolumn{7}{r}{Estimated NDR sum, maximum record per channel\quad} & $1.71520$\,Tbit/s \\
\bottomrule
\end{tabular}
\end{table}
\FloatBarrier
\section{Comb source and amplified-comb characterization}
\label{sec:comb}
The \SiN microresonator source chip (D230) was fabricated using a copper-contamination-free process~\cite{ji_deterministic_2025}. An external-cavity diode laser at $1551.213$\,nm, amplified to $800$\,mW in the input fibre, was tuned across the resonance into the effectively red-detuned regime to initiate a single dissipative Kerr soliton~\cite{herr_temporal_2014,ji_deterministic_2025} (main text Fig.~4a). A fibre Bragg grating suppressed the residual pump. The total resonator output power, including that pump, was $150$\,mW.

\begin{figure}[!htbp]
\centering
\includegraphics[width=\suppfigurewidth]{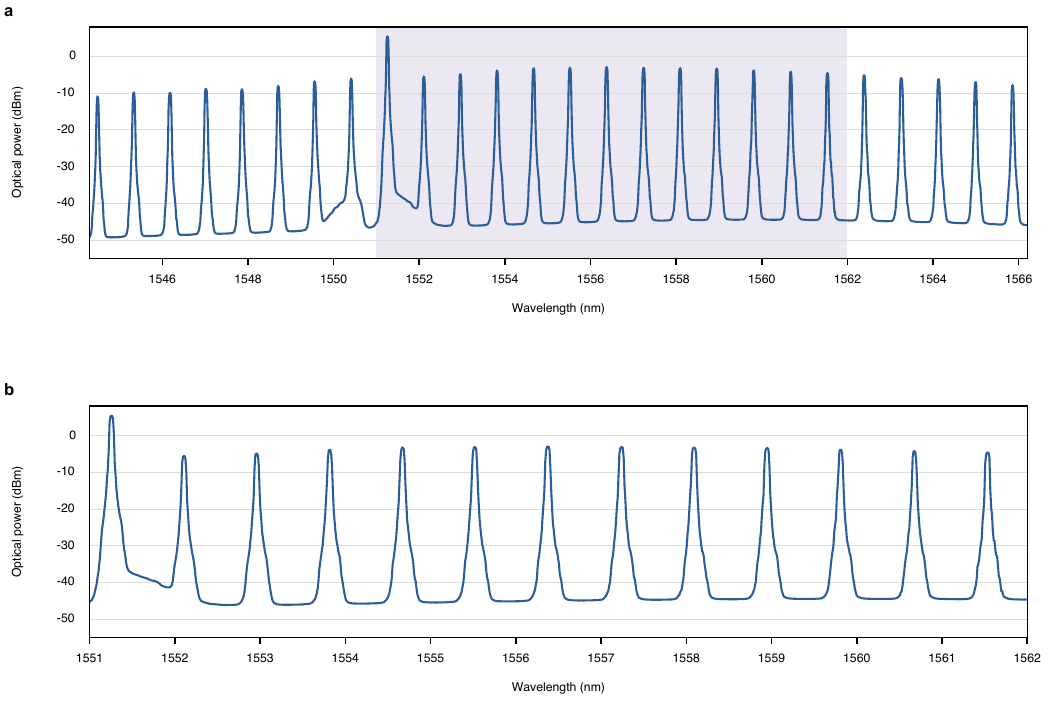}
\caption{\textbf{Amplified-comb spectrum.} \textbf{(a)} Overview, with shading marking the interval expanded in \textbf{(b)}. \textbf{(b)} The same trace showing individual comb peaks and the inter-line background. Both panels use the same power calibration.}
\label{fig:combspec}
\end{figure}

After spectral flattening and amplification, the selected carriers were characterized at the transmitter input using a high-resolution OSA with a $12.5$\,GHz resolution bandwidth. Their carrier grid differs from the laser-fed grid in Sec.~\ref{sec:tx}. The comb-line optical carrier-to-noise ratio (OCNR) is approximately $30$\,dB or lower. Figure~\ref{fig:combspec}a shows a separate amplified-comb spectrum, and Fig.~\ref{fig:combspec}b expands the shaded interval to resolve the individual peaks and inter-line background.

Table~\ref{tab:comb} lists the four comb-driven channel records corresponding to the transmission results in main text Fig.~4f.

\begin{table}[htbp]
\centering
\caption{\textbf{Comb-driven transmission at $80$\,GBd.} The four measured channels have PAM4-equivalent information rates. AIR and estimated NDR follow the definitions in Sec.~\ref{sec:tx}.}
\label{tab:comb}
\setlength{\tabcolsep}{5pt}
\begin{tabular}{lccccc}
\toprule
Channel & BER & CSNR (dB) & GMI (bit/sym) & AIR (Gbit/s) & NDR (Gbit/s) \\
\midrule
1 & $0.0035$ & $12.27$ & $1.960$ & $156.82$ & $148.90$ \\
2 & $0.0103$ & $10.83$ & $1.893$ & $151.47$ & $144.30$ \\
3 & $0.0165$ & $10.06$ & $1.831$ & $146.52$ & $139.73$ \\
4 & $0.0026$ & $12.34$ & $1.973$ & $157.86$ & $149.81$ \\
\midrule
\multicolumn{5}{r}{Estimated NDR sum, four measured channels} & $582.74$\,Gbit/s \\
\bottomrule
\end{tabular}
\end{table}

\FloatBarrier

\section{Carrier power and optical noise in comb-driven transmission}
\label{sec:ocnr}
To examine how carrier power and optical noise affect the comb-driven transmission results in Sec.~\ref{sec:comb}, we consider a simple OCNR model and a measured laser-carrier-power sweep.

\subsection{OCNR of an amplified comb line}
Consider a comb line with power $P_0$ at the input of a single EDFA. Let $G$ and $F$ be the amplifier's linear power gain and noise figure, respectively, and assume negligible input ASE. Using a two-polarization ASE convention, the OCNR in a reference optical bandwidth $B_{\mathrm{ref}}$ is
\begin{equation}
\mathrm{OCNR} = \frac{G\,P_0}{h\nu\,B_{\mathrm{ref}}\,(F G - 1)}
\;\xrightarrow[\,G\gg 1\,]{}\;
\frac{P_0}{h\nu\,B_{\mathrm{ref}}\,F},
\label{eq:ocnr}
\end{equation}
where $h$ is Planck's constant, $\nu$ is the optical frequency and $B_{\mathrm{ref}}=12.5$\,GHz. In the high-gain limit, OCNR is determined by the input line power and amplifier noise figure. Here, $P_0$ refers to the line power before amplification. The comb-line power range used for comparison with the carrier-power sweep was measured after amplification at the transmitter input.

\subsection{Effect on the received signal}
In a direct-detection receiver, beating between the modulated optical signal and ASE produces a signal-dependent electrical noise term. When signal--ASE beating dominates and the modulation depth and optical and electrical filtering are fixed, electrical SNR scales with OCNR. The recovered CSNR also includes contributions from receiver noise, ASE--ASE beating and filtering.

\subsection{Measured signal quality versus carrier power}
\label{sec:ocnrsweep}
\begin{figure}[!htbp]
\centering
\includegraphics[width=\suppfigurewidth]{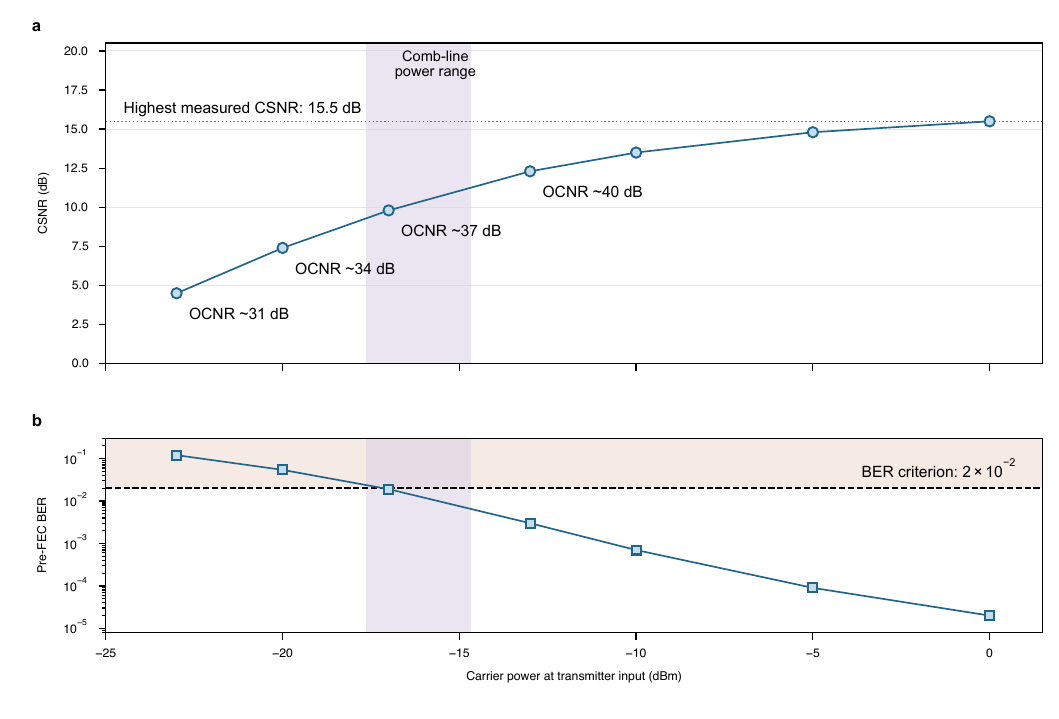}
\caption{\textbf{Signal quality versus laser-carrier power at $80$\,GBd PAM4.} \textbf{(a)} CSNR, with lower-power OCNR readings annotated and the highest measured CSNR ($15.5$\,dB) marked by a dotted line. Shading marks the comb-line power range ($-17.65$ to $-14.66$\,dBm). \textbf{(b)} Corresponding pre-FEC BER, with the $2\times10^{-2}$ reference criterion shown by a dashed line. Connecting lines guide the eye.}
\label{fig:ocnrsweep}
\end{figure}

An attenuated continuous-wave laser carrier was used to measure single-channel signal quality versus transmitter-input power at $80$\,GBd PAM4 (Table~\ref{tab:powersweep}). Over the measured range, CSNR increases from $4.5$\,dB at $-23$\,dBm to $15.5$\,dB at $0$\,dBm (Fig.~\ref{fig:ocnrsweep}a), while BER decreases from $1.2\times10^{-1}$ to $2\times10^{-5}$ (Fig.~\ref{fig:ocnrsweep}b). At $-17$\,dBm, BER is $1.9\times10^{-2}$, close to the adopted $2\times10^{-2}$ reference criterion. This input power lies within the measured comb-line range, which is shaded in Fig.~\ref{fig:ocnrsweep}a.

\begin{table}[htbp]
\centering
\caption{\textbf{Single-channel laser-fed carrier-power sweep at $80$\,GBd PAM4.} $P_{\mathrm{in}}$ is the carrier power at the transmitter input. OCNR values use a $12.5$\,GHz reference bandwidth. The adopted pre-FEC BER reference criterion is $2\times10^{-2}$.}
\label{tab:powersweep}
\begin{tabular}{lccccccc}
\toprule
$P_{\mathrm{in}}$ (dBm) & $0$ & $-5$ & $-10$ & $-13$ & $-17$ & $-20$ & $-23$ \\
\midrule
OCNR (dB) & $>40$ & $>40$ & $>40$ & $\sim$40 & $\sim$37 & $\sim$34 & $\sim$31 \\
CSNR (dB) & $15.5$ & $14.8$ & $13.5$ & $12.3$ & $9.8$ & $7.4$ & $4.5$ \\
BER & $2\times10^{-5}$ & $9\times10^{-5}$ & $7\times10^{-4}$ & $3\times10^{-3}$ & $1.9\times10^{-2}$ & $5.4\times10^{-2}$ & $1.2\times10^{-1}$ \\
\bottomrule
\end{tabular}
\end{table}

\FloatBarrier

\clearpage

\section{Channel-selective modulation and adjacent-pair transmission}
\label{sec:crosstalk}
\begin{figure}[!htbp]
\centering
\includegraphics[width=\suppfigurewidth]{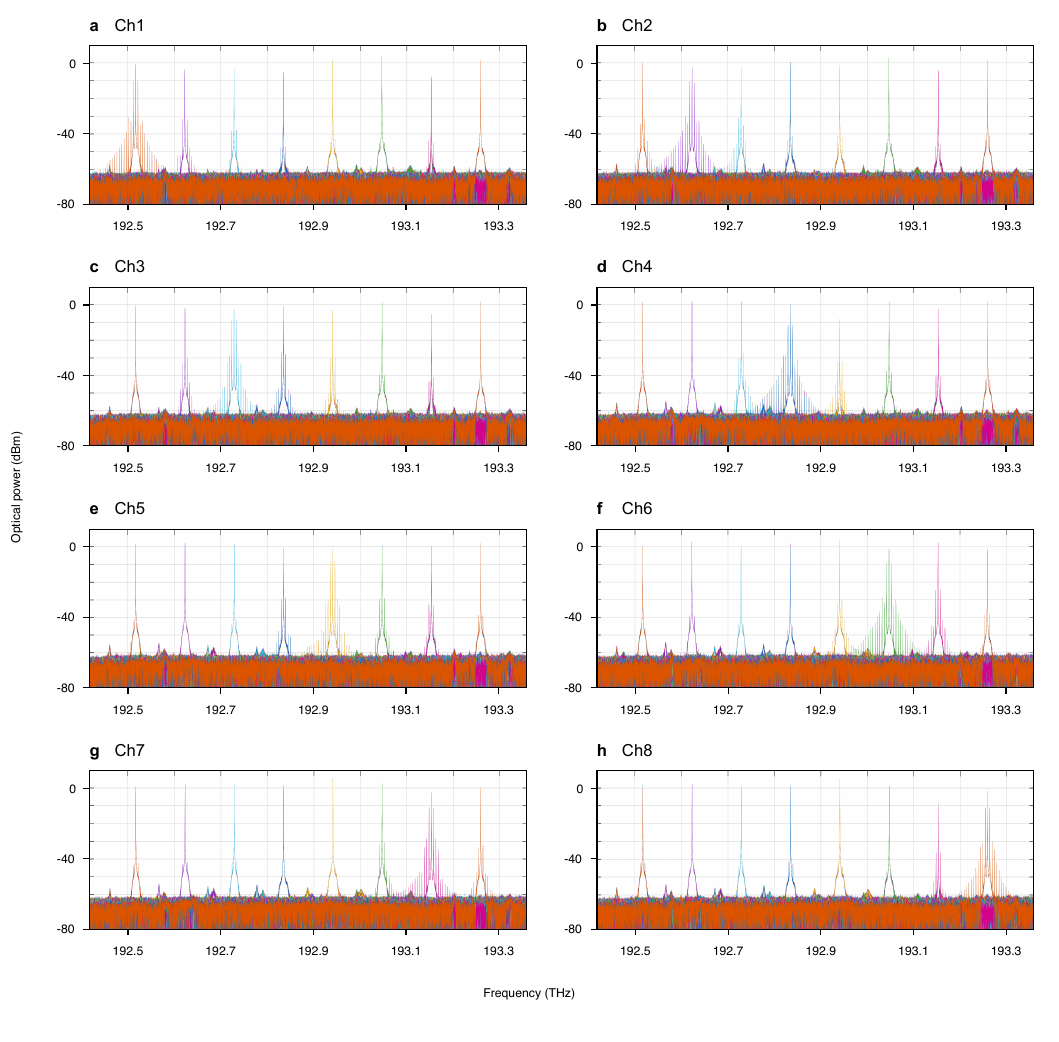}
\caption{\textbf{Common-output spectra for channel-selective modulation.} Panels \textbf{(a--h)} show the overlaid traces for addressed channels Ch1--Ch8, respectively, recorded with a $150$\,MHz resolution bandwidth. Spectral broadening identifies the driven carrier in each panel.}
\label{fig:crosstalk}
\end{figure}

The adjacent-pair transmission measurements in Sec.~\ref{sec:tx} are complemented by spectra recorded at the common transmitter output as each channel is addressed in turn (Fig.~\ref{fig:crosstalk}a--h). Spectral broadening identifies the driven carrier within the multiplexed grid.

\FloatBarrier

\clearpage
\bibliographystyle{apsrev4-2}
\bibliography{main_refs}